\documentclass[11pt]{article}
\pdfoutput=1
\usepackage{amsmath,amsfonts,amssymb,color}
\usepackage{cite,url}
\usepackage{graphicx}
\usepackage[utf8]{inputenc} 
\usepackage[bottom]{footmisc}
\usepackage{hyperref}

\allowdisplaybreaks

\def\msk2lam{$m_s\approx 5.5$~TeV, $m_a\approx 4.2$~TeV}

\def\smallCDF{{\rm{\scriptscriptstyle CDF}}}
\def\smallSM{{\rm{\scriptscriptstyle SM}}}
\def\smallBSM{{\rm{\scriptscriptstyle BSM}}}
\def\smallTSM{{\rm{\scriptscriptstyle TSM}}}

\def\smallH{{\scriptscriptstyle H}}

\def\smallR{{\scriptscriptstyle R}}
\def\smallT{{\scriptscriptstyle T}}

\def\beq{\begin{equation}}
\def\eeq{\end{equation}}
\def\bea{\begin{eqnarray}}
\def\eea{\end{eqnarray}}
\def\nn{\nonumber}

\def\aem{\alpha_{\rm em}}
\def\mHq{m^2_H}

\def\mHpq{m^2_{H^\pm}}

\def\TSM{${\rm TSM}$}

\def\sq2{\sqrt{2}}

\def\msbar{\overline{\rm MS}}

\def\smallmsbar{\scriptscriptstyle{\overline{\rm MS}}}

\def\smallIPI{{\scriptscriptstyle {\rm 1PI}}}

\long\def\symbolfootnote[#1]#2{\begingroup%
\def\thefootnote{\fnsymbol{footnote}}\footnote[#1]{#2}\endgroup}

\makeatletter
\newcommand{\vast}{\bBigg@{3}}
\makeatother

\begin{document}

\begin{titlepage}

\begin{center}

\vspace{1cm}

{\LARGE \bf On the two-loop BSM corrections to $h\longrightarrow
  \gamma\gamma$ }\\[3mm]

{\LARGE \bf in a triplet extension of the SM} 

\vspace{1cm}

{\Large Giuseppe Degrassi$^{\,a}$ and Pietro~Slavich$^{\,b}$}

\vspace*{1cm}

{\sl ${}^a$ 
Dipartimento di Matematica e Fisica, Università di Roma Tre, 

and INFN, Sezione di Roma Tre, I-00146 Rome, Italy.}
\vspace*{2mm}\\{\sl ${}^b$
   Sorbonne Université, CNRS,
  Laboratoire de Physique Th\'eorique et Hautes Energies, 
 
  LPTHE, F-75005, Paris, France.}
\end{center}
\symbolfootnote[0]{{\tt e-mail:}}
\symbolfootnote[0]{{\tt giuseppe.degrassi@uniroma3.it}}
\symbolfootnote[0]{{\tt slavich@lpthe.jussieu.fr}}

\vspace{0.7cm}

\abstract{We compute the two-loop BSM contributions to the
  $h\longrightarrow \gamma\gamma$ decay width in the SM extended with
  a real triplet of $SU(2)$. We consider scenarios in which the
  neutral components of doublet and triplet do not mix, so that the
  lighter neutral scalar $h$ has (at least approximately) SM-like
  couplings to fermions and gauge bosons. We focus on the two-loop
  corrections controlled by the quartic scalar couplings, and obtain
  explicit and compact formulas for the $h \gamma \gamma$ amplitude by
  means of a low-energy theorem that connects it to the derivative of
  the photon self-energy w.r.t.~the Higgs field.  We briefly discuss
  the numerical impact of the newly-computed contributions, showing
  that they may be required for a precise determination of
  $\Gamma[h\rightarrow \gamma \gamma]$ in scenarios where the quartic
  scalar couplings are large.}

\vfill

\end{titlepage}


\setcounter{footnote}{0}
\vspace*{-1.4cm}
\section{Introduction}
\label{sec:intro}
The discovery of a Higgs boson with mass around
$125$~GeV~\cite{CMS:2012qbp, ATLAS:2012yve} and properties compatible
with the predictions of the Standard Model
(SM)~\cite{ParticleDataGroup:2022pth} goes a long way towards
elucidating the mechanism of electroweak (EW) symmetry breaking, but
does not by itself preclude the existence of additional Higgs bosons
with masses around or even below the TeV scale, which could still be
discovered in the current or future runs of the LHC. Apart from the
requirement that they allow for a $125$-GeV scalar with SM-like
couplings to fermions and gauge bosons, the parameters of
beyond-the-SM (BSM) models with an extended Higgs sector are subject
to a number of experimental constraints from direct searches for new
particles, EW precision observables, and flavor physics, as well as
theory-driven constraints from perturbative unitarity, perturbativity,
and the stability of the scalar potential. However, it is often found
that, in BSM models, some of the quartic Higgs couplings can be of
${\cal O}(1$--$10)$ without violating any of the above-mentioned
constraints. This applies in particular to ``bottom-up'' constructions
in which the extended Higgs sector is not embedded in a supersymmetric
model supposed to be valid up to scales much higher than the
TeV. Couplings in this range may induce sizable radiative corrections
to the predictions for physical observables, up to the point where one
might wonder whether, in any given calculation, the uncomputed
higher-order effects spoil the accuracy of the prediction.

In the Two-Higgs-Doublet Model
(THDM)~\cite{Gunion:1989we,Aoki:2009ha,Branco:2011iw}, one of the
simplest and best-studied extensions of the SM, the viability of
scenarios with large quartic Higgs couplings motivated a number of
recent studies in which radiative corrections involving those
couplings were computed at the two-loop level. In particular, the
two-loop corrections to the $\rho$ parameter were computed in
refs.~\cite{Hessenberger:2016atw, Hessenberger:2022tcx}, various
effects of the two-loop corrections to the scalar mass matrices were
examined in ref.~\cite{Braathen:2017izn}, the two-loop corrections to
the trilinear self-coupling of the SM-like Higgs boson,
$\lambda_{hhh}$, were computed in refs.~\cite{Braathen:2019pxr,
  Braathen:2019zoh}, and the two-loop corrections to the decay width
for the process $h\longrightarrow \gamma \gamma$ were computed in
refs.~\cite{Degrassi:2023eii, Aiko:2023nqj}. In all cases it was found
that the two-loop corrections can significantly modify the one-loop
predictions, and should be taken into account for a precise
determination of the considered observable.

In this paper we consider an extension of the SM, which we refer to as
the TSM, where the Higgs sector is augmented with a real ($Y=0$)
triplet of $SU(2)$, so that the spectrum of physical Higgs states
includes two neutral scalars, $h$ and $H$, and a charged scalar,
$H^\pm$. The model was first introduced in ref.~\cite{Ross:1975fq},
and the contributions of the triplet to the EW precision observables
were computed in refs.~\cite{Passarino:1989py, Lynn:1990zk,
  Blank:1997qa, Forshaw:2001xq, Chen:2005jx, Chen:2006pb,
  Chankowski:2006hs, Chivukula:2007koj, Chen:2008jg}, where different
ways to connect the parameters of the TSM to a set of physical inputs
were extensively discussed. Various aspects of the phenomenology of
the TSM at the LHC were explored in refs.~\cite{FileviezPerez:2008bj,
  Wang:2013jba, Chabab:2018ert, Bell:2020gug, Chiang:2020rcv,
  Ashanujjaman:2023etj, Butterworth:2023rnw, Ashanujjaman:2024pky,
  Crivellin:2024uhc}, and the possibility that the recent measurement
of the $W$ boson mass by CDF~\cite{CDF:2022hxs}, which is $7\sigma$
away from the SM prediction\,\footnote{Note that more-recent
measurements of $m_W$ by ATLAS~\cite{ATLAS:2024erm} and
CMS~\cite{CMS:2024nau} are within $1\sigma$ from the SM prediction.},
is accommodated in the TSM through a small vacuum expectation value
(vev) for the triplet was discussed, e.g., in
refs.~\cite{FileviezPerez:2022lxp, Senjanovic:2022zwy}.

Any realistic extension of the SM must be able to accommodate the
measured -- and essentially SM-like -- properties of the Higgs boson
already discovered at the LHC. Among those properties, the width for
the loop-induced decay $h\longrightarrow \gamma \gamma$ plays a
special role, because the corresponding signal strength is already
measured with an accuracy of about
$6\%$~\cite{ParticleDataGroup:2022pth}, bound to further improve as
more data are collected at the LHC, and because both SM and BSM
particles contribute to it at the leading order (LO), i.e., at the
one-loop level. Indeed, most of the phenomenological analyses of the
TSM in refs.~\cite{FileviezPerez:2008bj, Wang:2013jba, Chabab:2018ert,
  Bell:2020gug, Chiang:2020rcv, Ashanujjaman:2023etj,
  Butterworth:2023rnw, Ashanujjaman:2024pky,
  Crivellin:2024uhc} considered the constraints on the model's
parameter space that arise from the LO prediction for $\Gamma(
h\rightarrow \gamma \gamma)$, which receives a BSM contribution from
one-loop diagrams involving the charged scalar $H^\pm$. In this paper
we take one step beyond the LO, and compute the BSM contributions to
$\Gamma( h\rightarrow \gamma \gamma)$ from two-loop diagrams
controlled by the scalar couplings involving the triplet, which, as
mentioned earlier, can in principle be of ${\cal O}(1$--$10)$. We
closely follow our approach from ref.~\cite{Degrassi:2023eii}, where
an analogous calculation was performed for the THDM. In particular, we
consider only ``aligned'' scenarios in which the neutral components of
doublet and triplet do not mix, so that the lighter neutral scalar
$h$, which we assume to be part of the doublet, has (at least
approximately) SM-like couplings to fermions and gauge bosons. The
choice of focusing on the effects of the potentially-large scalar
couplings that involve the triplet allows us to neglect the
contributions of two-loop diagrams involving the gauge and Yukawa
couplings (except of course for the electromagnetic couplings of the
external photons), and to neglect the mass of the SM-like Higgs boson
w.r.t.~the masses of the triplet-like states. In this case, we can
exploit a low-energy theorem (LET) that connects the $h \gamma \gamma$
amplitude to the derivative of the photon self-energy w.r.t.~the vev
of the SM-like Higgs field~\cite{Shifman:1979eb, Kniehl:1995tn}, but
we also cross-check our results via a direct calculation of the
amplitude.

\bigskip

The rest of the article is organized as follows: in
section~\ref{sec:tree} we fix our notation for the Higgs sector of the
TSM and introduce two distinct scenarios in which the neutral
components of doublet and triplet do not mix; in
section~\ref{sec:twoloop} we outline our calculation of the dominant
two-loop corrections to the decay width for $h\longrightarrow \gamma
\gamma$ in the two considered scenarios; in section~\ref{sec:numerics}
we briefly discuss the numerical impact of the newly-computed
corrections; section~\ref{sec:conclusions} contains our conclusions;
finally, three appendices collect explicit formulas for the one-loop
self-energies and tadpoles that are relevant to our calculation, for
the two-loop BSM contribution to the $h\gamma\gamma$ amplitude, and
for the Higgs-mass dependence of the $h\gamma\gamma$ amplitude at one
loop.

\section{The Higgs sector of the \TSM}
\label{sec:tree}

In this section we describe the Higgs sector of the \TSM\ at the tree
level. The scalar potential of the SM extended with an $SU(2)$ triplet
with $Y=0$ can be parametrized as\,\footnote{ We employ the same
conventions as in the ``{\tt SM+Triplet-Real}'' model file implemented
in the code {\tt SARAH}~\cite{Staub:2008uz, Staub:2009bi,
  Staub:2010jh, Staub:2012pb, Staub:2013tta}. For comparison, the
conventions of ref.~\cite{Bell:2020gug} correspond to the
replacements~
$
\mu^2\rightarrow -\mu_H^2\,,\,
m_T^2\rightarrow  - \mu_\Sigma^2\,,\,
\lambda_H \rightarrow 2\,\lambda_H\,,\,
\lambda_T\rightarrow  b_4\,,\,
\kappa_{HT}\rightarrow  a_1/\sq2\,,\,
\lambda_{HT}\rightarrow  a_2\,,\,
v\rightarrow  v_H\,,\,
v_T\rightarrow  \sq2\,v_{\Sigma}
\,$.
For an early discussion of the tree-level scalar potential and mass
spectrum of the TSM, see ref.~\cite{Chardonnet:1993wd}.  }
%
\bea
V_0 &=&
\mu^2\,H^{\dagger}H ~+~ \frac{m_T^2}2\,{\rm Tr}(T^2)
~+~\frac{\lambda_H}{2}\,(H^{\dagger}H)^2~+~\frac{\lambda_T}2\,{\rm Tr}(T^4)\nn\\
&&+~\kappa_{HT}\,H^{\dagger}T H
~+~\frac{\lambda_{HT}}2\,H^{\dagger}H\,{\rm Tr}(T^2)~,
\label{eq:V0}
\eea
where $H$ is the SM-like Higgs doublet and $T$ is the triplet. The
scalar potential in eq.~(\ref{eq:V0}) is bounded from below if
$\lambda_H>0$, $\lambda_T>0$ and $\lambda_{HT} >
-\sqrt{2\lambda_H\lambda_T}$. Expanded around their vevs $v$ and
$v_T$, respectively, the doublet and triplet fields can be decomposed
as
\beq
H ~=~ \frac{1}{\sqrt2}\left(\!\begin{array}{c} \sq2\,\phi_H^+ \\[2mm]
v\,+\,\phi^0_H\,+\,i\,G^0
\end{array}\!\right)~,~~~~~
T ~=~ \frac{1}{\sqrt2} \left(\!\begin{array}{cc} \frac{v_T}{\sq2} \,+\,\phi^0_T~
&\sq2\,\phi_T^+ \\[2mm] 
\sq2\,\phi_T^-&-\frac{v_T}{\sq2} \,-\,\phi^0_T  \end{array}\!\right)~.
\eeq
The tree-level tadpoles for the doublet and the triplet, respectively,
are
\bea
\label{eq:min1}
\tau_{\smallH} \equiv \left.\frac{\partial V_0}{\partial \phi^0_H}\right|_{\rm min}&=&
v\,\left(\mu^2 + \frac{\lambda_H}2\,v^2 + \frac{\lambda_{HT}}4\,v_T^2
- \frac{\kappa_{HT}}2\,v_T\right),\\[2mm]
\label{eq:min2}
\tau_{\smallT} \equiv \left.\frac{\partial V_0}{\partial \phi^0_T}\right|_{\rm min}&=&
\frac{v_T}{\sq2}\,\left(m_T^2 + \frac{\lambda_{T}}2\,v_T^2
 + \frac{\lambda_{HT}}2\,v^2 - \frac{\kappa_{HT}}2\frac{v^2}{v_T}\right).
\eea
The tree-level minimum conditions for the scalar potential correspond
to $\tau_{\smallH}=\tau_{\smallT}=0$\,, and a realistic EWSB scenario
requires $v\neq0$. As discussed, e.g., in
ref.~\cite{FileviezPerez:2008bj}, when $\kappa_{HT} = 0$ the only
realistic solution of the minimum conditions leads to $v^2 =
-2\,\mu^2/\lambda_H$ and $v_T=0$ (another possible solution where both
$v$ and $v_T$ are different from zero involves a massless physical
scalar).  Conversely, eq.~(\ref{eq:min2}) shows that $\kappa_{HT} \neq
0$ implies in turn $v_T \neq 0$\,.
In the latter case, the minimum conditions can be exploited to express
both of the mass parameters $\mu^2$ and $m_T^2$ as combinations of the
other Lagrangian parameters and the vevs. The tree-level mass matrices
${\cal M}^2_0$ and ${\cal M}^2_\pm$ for the neutral scalars
$(\phi^0_H,\phi^0_T)$ and the charged scalars
$(\phi^\pm_H,\phi^\pm_T)$, respectively, can thus be written as
\beq
\label{eq:mass0}
{\cal M}^2_0 ~=~
\left(\!\begin{array}{cc}
\lambda_H\,v^2 
& \frac{v}{\sq2}\,(\lambda_{HT}\,v_T-\kappa_{HT})\\[2mm]
\frac{v}{\sq2}\,(\lambda_{HT}\,v_T-\kappa_{HT})~~&
\lambda_T\,v_T^2 + \frac{\kappa_{HT}}2\frac{v^2}{v_T} 
\end{array}\!\right)~,
\eeq
\beq
\label{eq:masspm}
{\cal M}^2_\pm ~=~
\left(\!\begin{array}{cc}
\kappa_{HT}\,v_T 
& \frac{\kappa_{HT}}{\sq2}\,v\\[2mm]
\frac{\kappa_{HT}}{\sq2}\,v&
\frac{\kappa_{HT}}2\frac{v^2}{v_T} 
\end{array}\!\right)~.
\eeq
The mass eigenstates are obtained via rotations by the
angles\,\footnote{We will henceforth use the shortcuts $c_\theta\equiv
\cos\theta$, $s_\theta\equiv \sin\theta$, and
$t_\theta\equiv\tan\theta$ for a generic angle $\theta$.}  $\gamma$
and $\delta$:
\beq
\label{eq:rotate}
\left(\!\begin{array}{c} h \\ H \end{array}\!\right) ~=~
\left(\!\begin{array}{rr} c_\gamma&\!s_\gamma\\-s_\gamma
&\! c_\gamma \end{array}\!\right)\, \left(\!\begin{array}{c}
\phi^0_H\\ \phi^0_T \end{array}\!\right)~,~~~~~
\left(\!\begin{array}{c} G^\pm \\ H^\pm \end{array}\!\right) ~=~
\left(\!\begin{array}{rr} c_\delta&\!s_\delta\\-s_\delta
&\! c_\delta \end{array}\!\right)\,\left(\!\begin{array}{c}
\phi_H^\pm\\ \phi_T^\pm \end{array}\!\right)~.
\eeq
In particular, for the charged scalars one finds
\beq
\label{eq:charged}
\delta ~=~ -\arctan\left(\frac{\sq2\,v_T}v\right)~,~~~~~~
m^2_{G^\pm}~=~0,~~~~~~
m^2_{H^\pm}~=~\kappa_{HT}\,v_T\,\left(1+\frac{v^2}{2\,v_T^2}\right)~.
\eeq
Finally, the tree-level masses for the gauge bosons and the tree-level
Fermi constant are
\beq
\label{eq:gauge}
m_Z^2~=~ \frac{g^2 + g^{\prime\,2}}{4}\,v^2~,~~~~~
m_W^2~=~ \frac{g^2}{4}\,(v^2 + 2\,v_T^2) ~=~ \frac{g^2}{4}\,v^2\,c_\delta^{-2}~,~~~~~
(\sq2\,G_\mu)^{-1} ~=~ (v^2 + 2\,v_T^2) ~=~ v^2\,c_\delta^{-2}~,
\eeq
where $g$ and $g^\prime$ are the $SU(2)$ and $U(1)$ gauge couplings,
respectively. We recall here that the usual approach to the analysis
of the EW sector of the SM consists in treating $m_Z$, $G_\mu$, and
$\alpha_{\rm em}\equiv e^2/(4\pi)$, where $e$ is the electric charge,
as physical input quantities, and obtaining predictions for $m_W^2$
and for $\theta_W \equiv \arctan(g^\prime/g)$. In extensions of the
SM we can write:
\beq
\label{eq:rho}
m_W^2 ~=~ \frac12\,m_Z^2\,\rho \,
\left(1+\sqrt{1-\frac{2\,\sqrt2\,\pi\,\alpha_{\rm em}}{G_\mu \,m_Z^2\,\rho}}
\right)~,~~~~~
s_{\theta_W}^2 ~=~ \frac12\,
\left(1-\sqrt{1-\frac{2\,\sqrt2\,\pi\,\alpha_{\rm em}}{G_\mu \,m_Z^2\,\rho}}
\right)~,
\eeq
where $\rho \equiv m_W^2/(m_Z^2\,c_{\theta_W}^2)$ is treated as an
additional input parameter. In the SM $\rho=1$ at the tree level, but
in the TSM the presence of a (small) vev for the triplet leads to
$\rho\gtrsim1$. As discussed, e.g., in ref.~\cite{Benakli:2022gjn},
this causes shifts in both $m_W^2$ and $\theta_W$ w.r.t.~the
corresponding SM predictions.  Expanding eq.~(\ref{eq:rho}) in
$\Delta\rho \equiv \rho-1$, we get:
\beq
\Delta m_W^2 ~\approx~
\frac{c_{\theta_W}^2}{c_{\theta_W}^2-s_{\theta_W}^2}\,m_W^2\,\Delta\rho~,~~~~
\Delta s_{\theta_W}^2 ~\approx~
-\frac{c_{\theta_W}^2s_{\theta_W}^2}{c_{\theta_W}^2-s_{\theta_W}^2}\,\Delta\rho~.
\eeq
Eq.~(\ref{eq:gauge}) shows that in the TSM $\rho = c_\delta^{-2}$ at
the tree level, hence $\Delta\rho = t_\delta^2 = 2\,
v_T^2/v^2$. Reconciling the SM prediction
$m_W^\smallSM=80.356$~GeV~\cite{ParticleDataGroup:2022pth} with the
CDF measurement $m_W^\smallCDF=80.433$~GeV~\cite{CDF:2022hxs}
would require $v_T\approx 6.4$~GeV.\footnote{Neglecting the effect of
the triplet vev on the prediction for $\theta_W$, as done, e.g., in
refs.~\cite{FileviezPerez:2022lxp, Senjanovic:2022zwy,
  Ashanujjaman:2023etj}, one would instead find $\Delta m_W^2 \approx
m_W^2\,\Delta\rho$\,, and the CDF result would imply $v_T\approx
7.6$~GeV with our normalization of the vev.}

\bigskip

In our calculation of the diphoton decay width of the SM-like Higgs
boson we will focus on scenarios in which the neutral components of
doublet and triplet do not mix, i.e., $\gamma=0$. According to
eq.~(\ref{eq:mass0}), this can be realized when $\kappa_{HT} =
\lambda_{HT}\,v_T$, in which case the minimum condition
$\tau_{\smallT}=0$ implies $m_T^2 = - \lambda_T\,v_T^2/2$, see
eq.~(\ref{eq:min2}), and the tree-level masses of the physical states
can be written as
\beq
\label{eq:massesnomix}
m_h^2~=~ \lambda_H\,v^2~,~~~~~~
m_H^2~=~ \frac{\lambda_{HT}}2\,v^2 + \lambda_{T}\,v_T^2~,~~~~~~
m_{H^\pm}^2~=~\frac{\lambda_{HT}}2\,v^2 \,\left(1+\frac{2\,v_T^2}{v^2}\right)~.
\eeq
In view of the small value of $v_T$, the neutral and charged
triplet-like states are almost degenerate in mass.  This scenario --
which we will call {\bf ``scenario A''} in the following -- is called
``fermiophobic'' in ref.~\cite{Butterworth:2023rnw} because the
neutral component of the triplet, $H$, does not couple to fermions and
can have sizeable branching ratios to pairs of $W$ bosons or photons
(in contrast, $H^\pm$ can still decay to fermion pairs through its
small doublet component). The neutral component of the doublet, $h$,
has approximately SM-like couplings to fermions and gauge bosons at
the tree level. In fact, the couplings of $h$ to fermion pairs and to
$Z$-boson pairs are enhanced by $c_\delta^{-1}$ w.r.t.~the
corresponding SM predictions expressed as combinations of masses and
$G_\mu$, and the couplings of $h$ to $W$-boson pairs are suppressed by
$c_\delta$, but even the relatively large value $v_T\approx 6.4$~GeV,
which could be considered as an upper bound of the acceptable range
for the triplet vev, corresponds to $c_\delta \approx 0.999$.

\vfill
\newpage

A different scenario in which the neutral doublet-like and
triplet-like states do not mix -- which we will call {\bf ``scenario
  B''} in the following -- is realized when $\kappa_{HT}= v_T=0$, in
which case the scalar potential possesses a discrete $Z_2$ symmetry
for $T\rightarrow -T$ that forbids the mixing. In this scenario
eq.~(\ref{eq:min2}) cannot be used to replace $m_T^2$ with a
combination of the other Lagrangian parameters and the vevs, and the
tree-level masses of the physical states can be written as
\beq
\label{eq:massesZ2}
m_h^2~=~ \lambda_H\,v^2~,~~~~~~
m_H^2~=~ m_{H^\pm}^2~=~ m_T^2 + \frac{\lambda_{HT}}2\,v^2~.
\eeq
The requirement that the SM-like minimum of the scalar potential be
deeper than an unphysical minimum with $v=0$ and $v_T\neq 0$ implies
$m_T^2 > -\sqrt{\lambda_T/2}\;m_h\,v$\,. When the triplet mass is
treated as an input parameter instead of $m_T^2$, this translates to
$\lambda_{HT} < 2\,(\mHpq +\sqrt{\lambda_T/2}\;m_h\,v)/v^2$.  We
remark that the mass spectra of the scenarios A and B coincide when
the limit $v_T\rightarrow0$ is taken in eq.~(\ref{eq:massesnomix}) and
the limit $m_T^2\rightarrow 0$ is taken in eq.~(\ref{eq:massesZ2}).

The degeneracy of the tree-level masses of the triplet-like states in
the scenario B is lifted by small radiative corrections involving the
EW gauge bosons, so that $m_{H^\pm} \gtrsim m_H$. The neutral
component of the triplet, $H$, is stable and can serve as a Dark
Matter (DM) candidate~\cite{Cirelli:2005uq}, while the neutral
component of the doublet, $h$, has exactly SM-like couplings to
fermions and to gauge bosons at the tree level. The charged scalar
$H^\pm$ decays to $H\pi^\pm$, $H e^\pm \nu_e$ or $H \mu^\pm \nu_\mu$,
giving rise to a disappearing track at the LHC because the charged
particle in the final state is too soft to be detected. The interplay
between the constraints on this scenario from DM searches and from
direct searches at the LHC was discussed in
refs.~\cite{FileviezPerez:2008bj, Bell:2020gug, Chiang:2020rcv}.

\bigskip

While the measured value of the Higgs mass determines the quartic
self-coupling of the doublet to be $\lambda_H\approx 0.26$ when
$\gamma=0$, the quartic couplings that involve the triplet,
$\lambda_{HT}$ and $\lambda_T$, are free parameters of the TSM,
bounded from above only by theoretical considerations. The upper
bounds from tree-level perturbative unitarity (i.e., the requirement
that the tree-level $2\rightarrow2$ scattering amplitudes remain
unitary at high energies) were discussed, e.g., in
refs.~\cite{Khan:2016sxm, Chabab:2018ert, Bell:2020gug}, and amount to
$\lambda_{HT}\lesssim 14$ and $\lambda_T \lesssim 5$. The upper bounds
from perturbativity (i.e., the requirement that the couplings
themselves are not so large as to spoil the convergence of the
perturbative series) involve some degree of subjectivity: for example,
some studies in the literature apply upper bounds of
$4\pi$~\cite{Forshaw:2003kh, Khan:2016sxm} or of
$2\sqrt\pi$~\cite{FileviezPerez:2008bj,
  Wang:2013jba,Butterworth:2023rnw} to both couplings, while
ref.~\cite{Bell:2020gug} applies upper bounds derived from the fixed
points of the two-loop RGEs,\footnote{The bound $\lambda_T \lesssim 2$
quoted in ref.~\cite{Bell:2020gug} was obtained with an older version
of the code {\tt SARAH}, in which the model file ``{\tt
  SM+Triplet-Real}'' contained a bug that affected the RGEs.}  leading
to $\lambda_{HT}\lesssim 8$ and $\lambda_T \lesssim 4$. In any case,
it appears that the couplings that involve the triplet can be
considerably larger than the SM couplings without violating any of the
theoretical bounds. In such scenarios, the radiative corrections
controlled by those couplings can be legitimately expected to be the
dominant ones.

\vfill
\newpage

\section{Leading two-loop contributions to
  $\Gamma[h\rightarrow \gamma \gamma]$}
\label{sec:twoloop}

We now discuss our calculation of the dominant two-loop corrections to
$\Gamma[h\rightarrow \gamma \gamma]$ in the TSM, in which we closely
follow the approach of the THDM calculation of
ref.~\cite{Degrassi:2023eii}. We consider the two scenarios mentioned
earlier, in which the neutral components of the doublet and triplet
scalars do not mix, and we focus on the two-loop BSM corrections that
are controlled by the potentially large quartic couplings
$\lambda_{HT}$ and $\lambda_T$. In contrast, we neglect all
corrections involving the gauge and Yukawa couplings. We also treat
the mass of the SM-like Higgs boson as negligible w.r.t.~the masses of
the BSM Higgs bosons, which allows us to use the LET of
refs.~\cite{Shifman:1979eb, Kniehl:1995tn} to connect the $h\gamma
\gamma$ amplitude to the derivative of the photon self-energy
w.r.t.~the doublet vev.

\subsection{The LET and the calculation of the photon self-energy}

The partial width for the $h\longrightarrow \gamma \gamma$ decay can
be written as
\beq
\label{eq:width}
\Gamma(h\rightarrow \gamma\gamma) ~=~
\frac{G_\mu\,\aem^2 M_h^3}{128\,\sqrt2\,\pi^3}
\,\left|{\cal P}_h^{1\ell}+{\cal P}_h^{2\ell}\right|^2~,
\eeq
where we recall that the Fermi constant $G_\mu$ is proportional to
$c_\delta^2/v^2$ at the tree level. ${\cal P}_h^{1\ell}$ and ${\cal
  P}_h^{2\ell}$ denote the one- and two-loop $h\gamma\gamma$
amplitudes, respectively. The latter can be further decomposed as
\beq
\label{eq:splitP2l}
{\cal P}_h^{2\ell} ~=~ {\cal P}_h^{2\ell,\,\smallIPI} \,+\,
\Delta {\cal P}_h^{1\ell}\,+\,
\delta {\cal P}_h^{1\ell} \,+\,K_r\, {\cal P}_h^{1\ell}~,
\eeq
where: ${\cal P}_h^{2\ell,\,\smallIPI}$ denotes the genuine two-loop
part, in which we include the one-particle-irreducible (1PI)
contributions as well as the $\msbar$ counterterm contributions;
$\Delta {\cal P}_h^{1\ell}$ stems from the minimum conditions for the
scalar potential, and from the conditions of vanishing SM-like Higgs
mass and vanishing mixing between the neutral scalars; $\delta {\cal
  P}_h^{1\ell}$ stems from renormalization-scheme choices for the
parameters entering ${\cal P}_h^{1\ell}$ (in particular, $\delta{\cal
  P}_h^{1\ell}=0$ would correspond to all parameters being in the
$\msbar$ scheme); the additional correction factor $K_r$ accounts for
the diagonal WFR of the external Higgs field and for the
renormalization of $G_\mu$.

\bigskip

In the approximation of vanishing external momentum for the
$h\gamma\gamma$ amplitude (i.e., vanishing mass for the SM-like Higgs
boson), the LET of refs.~\cite{Shifman:1979eb, Kniehl:1995tn} allows
us to write\,\footnote{Compared with the analogous formulas for the
THDM, see eq.~(39) of ref.~\cite{Degrassi:2023eii}, the additional
factors $c_\delta^{-1}$ stem from the different relation between
$G_\mu$ and $v$ in the TSM.}
\beq
\label{eq:LET}
    {\cal P}_h^{1\ell} ~=~ \frac{2\pi\,v\,c_\delta^{-1}}{\aem}~\frac{d\,
      \Pi_{\gamma\gamma}^{1\ell} (0)}{dv}~,
    ~~~~~~~~~
    {\cal P}_h^{2\ell,\,\smallIPI} ~=~ \frac{2\pi\,v\,c_\delta^{-1}}{\aem}\,\frac{d\,
      \Pi_{\gamma\gamma}^{2\ell}(0)}{dv}~,
\eeq
where $\Pi_{\gamma\gamma}(0)$ denotes the transverse part of the
dimensionless self-energy of the photon at vanishing external
momentum. At the one-loop level, it reads
\bea
\Pi_{\gamma\gamma}^{1\ell}(0)&=&
\Pi_{\gamma\gamma}^{1\ell,\,H^\pm}(0)
~+~\Pi_{\gamma\gamma}^{1\ell,\,t}(0)
~+~\Pi_{\gamma\gamma}^{1\ell,\,W}(0)\nonumber\\[2mm]
&=&
\frac{\aem}{4\pi}\,\left( \frac 13 \,\ln  \frac{\,m_{H^\pm}^2}{Q^2}
~+~ \frac 43 \,Q_t^2\,N_c \,\ln \frac{m_{t}^2}{Q^2}
~-~ 7 \,\ln \frac{m_{W}^2}{Q^2} \,+\,\frac 23 ~ \right)~,
\label{eq:Pigaga1l}
\eea
where $Q$ is the renormalization scale, $N_c=3$ is a color factor,
$Q_t=2/3$ is the electric charge of the top quark, and we omitted all
other fermionic contributions because the corresponding contributions
to the $h\gamma\gamma$ amplitude are suppressed by the small fermion
masses. The contribution of the gauge sector,
$\Pi_{\gamma\gamma}^{1\ell,\,W}(0)$, is in fact gauge dependent, and
only when computed in the unitary gauge or in the background-field
gauge (or using the pinch technique) can it be directly connected to
${\cal P}_h^{1\ell}$ through eq.~(\ref{eq:LET}).

\bigskip

We computed the contributions to the transverse part of the photon
self-energy from two-loop diagrams that involve the BSM Higgs bosons,
which we denote as $\Pi_{\gamma\gamma}^{2\ell,\,\smallBSM}(0)$, by
feeding to {\tt FeynArts}~\cite{Hahn:2000kx} the model file for the
TSM produced by {\tt SARAH}~\cite{Staub:2008uz, Staub:2009bi,
  Staub:2010jh, Staub:2012pb, Staub:2013tta}. We fixed the masses and
couplings of the BSM Higgs bosons to the values that correspond to
either the scenario A, see eq.~(\ref{eq:massesnomix}), or the scenario
B, see eq.~(\ref{eq:massesZ2}).  We performed our calculation in the
unitary gauge, including also the contributions from diagrams that
involve gauge bosons together with the BSM Higgs bosons. We set the
mass of the SM-like Higgs boson to zero from the start, but we
introduced a fictitious photon mass to regularize some IR-divergent
integrals. Next, we Taylor-expanded the self-energy in powers of the
external momentum $p^2$: the zeroth-order term of the expansion
vanishes as a consequence of gauge invariance, while the first-order
term corresponds to $\Pi_{\gamma\gamma}^{2\ell,\,\smallBSM}(0)$. We
evaluated the two-loop vacuum integrals using the results of
ref.~\cite{Davydychev:1992mt}. Finally, we sent the fictitious photon
mass to zero and then took the ``gaugeless limit'' of vanishing EW
gauge couplings, except for an overall factor $\aem$ from the
couplings of the external photons. For what concerns the counterterm
contributions to $\Pi_{\gamma\gamma}^{2\ell,\,\smallBSM}(0)$, we
assumed that the charged-Higgs mass $m^2_{H^\pm}$ entering
$\Pi_{\gamma\gamma}^{1\ell}(0)$ in eq.~(\ref{eq:Pigaga1l}) is
expressed in the $\msbar$ scheme (the definitions of the top and $W$
masses are irrelevant under our approximations), and we fixed the
counterterm of the charged-Higgs mass to the divergent part of the
one-loop self-energy $\Pi_{H^+H^-}(m^2_{H^\pm})$.

\bigskip

After the computation of the two-loop BSM contributions to the photon
self-energy, the genuine two-loop part of the $h\gamma\gamma$
amplitude, denoted as ${\cal P}_h^{2\ell,\,\smallIPI}$ in
eq.~(\ref{eq:splitP2l}), can be straightforwardly obtained through the
LET of refs.~\cite{Shifman:1979eb, Kniehl:1995tn}, by taking the
derivative of $\Pi_{\gamma\gamma}^{2\ell,\,\smallBSM}(0)$ w.r.t.~the
doublet vev $v$ as in eq.~(\ref{eq:LET}). To this purpose, the masses
of the BSM Higgs bosons must be set to the tree-level expressions
given in eqs.~(\ref{eq:massesnomix}) and (\ref{eq:massesZ2}) for the
scenarios A and B, respectively. However, the calculation of the
remaining contributions in eq.~(\ref{eq:splitP2l}) differs according
to which of the two scenarios with vanishing doublet-triplet mixing is
considered, and within each scenario there are multiple options for
the sets of parameters that can be treated as input, and multiple
possible choices of renormalization conditions. This will be discussed
in detail in sections \ref{sec:scenA} and \ref{sec:scenB}. Finally, we
stress that in both scenarios we cross-checked our LET-based
calculation of ${\cal P}_h^{2\ell}$ with a direct calculation of the
$h\gamma\gamma$ amplitude at vanishing external momentum, performed
along the lines of the calculation of
$\Pi_{\gamma\gamma}^{2\ell,\,\smallBSM}(0)$ described above.

\bigskip
\newpage

\subsection{Calculation of the $h\gamma\gamma$ amplitude in the
  scenario A}
\label{sec:scenA}

In this scenario, the squared masses of the SM particles entering the
one-loop part of the photon self-energy, eq.~(\ref{eq:Pigaga1l}), can
be expressed as $m_t^2 = y_t^2 \,v^2/2$\,, and $m_W^2 =
g^2\,v^2\,c_\delta^{-2}/4$, see eq.~(\ref{eq:gauge}). Recalling that
$c_\delta^{-2} = 1 + 2\,v_T^2/v^2$, the combination of
eqs.~(\ref{eq:LET}) and (\ref{eq:Pigaga1l}) yields for the SM-like
part of the one-loop $h\gamma\gamma$ amplitude
\beq
    {\cal P}_h^{1\ell,\,\smallSM}~=~
    \frac43\,Q_t^2\,N_c\,c_\delta^{-1} \,-\, 7\,c_\delta~.
\label{eq:P1l-A}
\eeq
We remark that, differently from the case of the aligned THDM
discussed in ref.~\cite{Degrassi:2023eii}, it is not entirely
appropriate to speak of a ``SM part'' of the amplitude, because even
when $\gamma=0$ the couplings of $h$ to fermions and gauge bosons
differ from their SM counterparts by powers of $c_\delta$. However, as
mentioned earlier in section~\ref{sec:tree}, for realistic values of
$c_\delta$ the numerical impact of this difference is minimal. We
therefore focus our two-loop calculation on the contributions of
diagrams that involve the BSM scalars, and omit the contributions of
diagrams that involve only SM particles.

\bigskip

We now move on to the calculation of the BSM part of the one-loop
$h\gamma\gamma$ amplitude, which we denote as ${\cal
  P}_h^{1\ell,\,\smallBSM}$, and of the associated counterterm
contributions. For consistency with the two-loop calculation of the
amplitude, when applying the LET theorem of eq.~(\ref{eq:LET}) to the
one-loop photon self-energy in eq.~(\ref{eq:Pigaga1l}) we cannot just
use the tree-level expression for the charged-Higgs mass given in
eq.~(\ref{eq:massesnomix}). Indeed, our choice for the counterterm
contributions to $\Pi_{\gamma\gamma}^{2\ell,\,\smallBSM}(0)$ implies
that, in the one-loop part of the photon self-energy, we must use
for charged-Higgs mass the tree-level expression -- in terms of
$\msbar$-renormalized Lagrangian parameters -- that is valid {\em
  before} we impose the minimum conditions for the scalar potential,
the condition of vanishing mass for the SM-like Higgs boson, and the
condition of vanishing triplet-doublet mixing:
\bea
\widehat{m}^2_{H^\pm} &=& \frac12\,\left[~
  \mu^2 + m_T^2 +\frac12\,(\lambda_H+\lambda_{HT})\,v^2
  + \frac14\,(\lambda_{HT}+2\,\lambda_T)\,v_T^2
  + \frac12 \,\kappa_{HT}\,v_T\right.
  \nonumber\\[2mm]
  && ~~~~~+\,\left.\sqrt{
    2\,\kappa_{HT}^2\,v^2 +
    \left(\mu^2 - m_T^2 +\frac12\,(\lambda_H-\lambda_{HT})\,v^2
    + \frac14 \,(\lambda_{HT}-2\,\lambda_T)\,v_T^2
    + \frac12 \,\kappa_{HT}\,v_T\right)^2
  }~\right]~.  \nonumber\\
\label{eq:mHptree}
\eea
Taking the derivative w.r.t.~$v$ of the charged-Higgs contribution to
the one-loop self-energy of the photon, and then removing the
parameters $\mu^2$, $m_T^2$, $\lambda_H$ and $\kappa_{HT}$ by means of
the tree-level conditions $\tau_\smallH = \tau_\smallT = m_h^2 =
\left({\cal M}_0^2\right)_{12} = 0$ -- where $\tau_\smallH$ and
$\tau_\smallT$ are given in eqs.~(\ref{eq:min1}) and (\ref{eq:min2}),
respectively, $m_h^2 = \lambda_H\,v^2$, and $\left({\cal
  M}_0^2\right)_{12}$ is given in eq.~(\ref{eq:mass0}) -- would lead
to ${\cal P}_h^{1\ell,\,\smallBSM} =\, c_{\delta}/3$. However, in the
two-loop calculation of the $h\gamma\gamma$ amplitude we need to
ensure that all parameters entering the one-loop amplitude are
consistently renormalized at the one-loop level. In particular, we
require
\beq
\tau_\smallH \,+\, T_h \,=\,0~,~~~~~
\tau_\smallT \,+\, T_H \,=\,0~,~~~~~
m_h^2 \,-\, \frac{T_h}{v} \,+\,\Pi_{hh}(0) \,=\,0~,~~~~~
\left({\cal M}_0^2\right)_{12}\,+\, \Pi_{hH}(0) \,=\,0~,
\label{eq:renorm}
\eeq
where: $\tau_\smallH$, $\tau_\smallT$, $m_h^2$ and $\left({\cal
  M}_0^2\right)_{12}$ are expressed in terms of $\msbar$-renormalized
parameters; $T_h$ and $T_H$ are the finite parts of the one-loop
tadpole diagrams\,\footnote{Decomposing the effective potential as
$V_0 + \Delta V$, we also have $T_\varphi \,=\, d \Delta
V/d\varphi$. Note that in the limit of vanishing doublet-triplet
mixing, $\gamma=0$, we can identify directly $\phi^0_H$ with $h$ and
$\phi^0_T$ with $H$.} for the neutral scalars; $\Pi_{hh}(0)$ and
$\Pi_{hH}(0)$ are the finite parts of the $(1,1)$ and $(1,2)$ entries
of the one-loop self-energy matrix for the neutral scalars, evaluated
at vanishing external momentum. Explicit formulas for the one-loop
tadpoles and self-energies, under the approximations relevant to our
two-loop calculation, are listed in the appendix A. The first two
conditions in eq.~(\ref{eq:renorm}) above imply that $v$ and $v_T$ are
the vevs of the loop-corrected effective potential, the third
condition implies that the {\em pole} mass of the SM-like Higgs boson
is vanishing, while the fourth condition implies that the
doublet-triplet mixing term in the loop-corrected mass matrix for the
neutral scalars vanishes for the external momentum that is relevant to
the calculation of the $h\longrightarrow\gamma\gamma$ decay width. At
the next-to-leading order (NLO) in the loop expansion, we thus find
\beq
\frac{2\pi\,v\,c_\delta^{-1}}{\aem}~\frac{d\,
  \Pi_{\gamma\gamma}^{1\ell,\,H^\pm} (0)}{dv}
~=~ \frac{c_{\delta}}{3} ~+~ \Delta {\cal P}_h^{1\ell,\,\smallBSM}~,
\label{eq:P1lBSM}
\eeq
where
\beq
\Delta {\cal P}_h^{1\ell,\,\smallBSM}~=~ -\frac{s_{\delta}\,t_{\delta}}{6\,\mHpq}\,\left[
  \Pi_{hh}(0) \,-\, \left(3 + 2\,c_{2\delta}\right)\,\frac{T_h}{v}
  \,-\, 2\,t_{\delta}^{-3} \left( \Pi_{hH}(0) - 2\,c_{2\delta}\,\frac{T_H}{v}
  \right)\right]~.
\label{eq:shift}
\eeq
We can then set ${\cal P}_h^{1\ell,\,\smallBSM} \,=\,
{c_{\delta}}/{3}$, while $\Delta {\cal P}_h^{1\ell,\,\smallBSM}$
becomes part of ${\cal P}_h^{2\ell}$ as the second contribution in
eq.~(\ref{eq:splitP2l}).

\bigskip

The third contribution to the two-loop $h\gamma\gamma$ amplitude in
eq.~(\ref{eq:splitP2l}) comes from the choice of renormalization
scheme for the parameters entering the one-loop amplitude. In the
scenario~A, ${\cal P}_h^{1\ell}$ depends only on $c_\delta$, which is
in turn a combination of $v$ and $v_T$. The various possible ways to
associate the doublet and triplet vevs (together with the EW gauge
couplings) to a set of physical input parameters have been extensively
discussed in early studies of the EW precision observables in the TSM,
see~refs.\cite{Passarino:1989py, Lynn:1990zk, Blank:1997qa,
  Forshaw:2001xq, Chen:2005jx, Chen:2006pb, Chankowski:2006hs,
  Chivukula:2007koj, Chen:2008jg}. In this work we follow the approach
advocated in ref.~\cite{Chankowski:2006hs}, i.e., we treat the
measured value of the Fermi constant $G_\mu$ and the
$\msbar$-renormalized triplet vev $v_T$ as input parameters, while we
treat the doublet vev $v$ (and, consequently, $c_\delta$) as a derived
quantity. In practice, we exploit the relation
\beq
\label{eq:cdGmu}
c_\delta^2~=~1\,-\,2\,\sq2\,G_\mu\,v_T^2
\eeq
to obtain
\beq
\delta{\cal P}_h^{1\ell}~=~ -\frac{t_\delta^2}{2}\,
\left(\frac13\,c_\delta \,-\,\frac43\,Q_t^2\,N_c\,c_\delta^{-1} \,-\,7\,c_\delta
\right) \left(\frac{\delta G_\mu}{G_\mu} \,+\,2\,\frac{\delta v_T}{v_T}\right)~,
\eeq
where for a parameter $x$ in a generic renormalization scheme $R$ we
define $x^\smallR = x^{\smallmsbar} - \delta x$. Our choice of
renormalization conditions corresponds then to $\delta v_T=0$, and,
under our approximations for the two-loop calculation,
\beq
\frac{\delta G_\mu}{G_\mu}~=~ - \frac{\Pi_{WW}^\smallBSM(0)}{m_W^2}~,
\label{eq:dGmu}
\eeq
where $\Pi_{WW}^\smallBSM(0)$ denotes the finite part of the BSM
contribution to the $W$ self-energy at vanishing external momentum, an
explicit formula for which is given in the appendix~A.

\bigskip

The fourth contribution to the two-loop amplitude ${\cal P}_h^{2\ell}$
in eq.~(\ref{eq:splitP2l}), i.e., $K_r\,{\cal P}_h^{1\ell}$, arises
from the diagonal wave-function renormalization (WFR) of the external
Higgs field and from the renormalization of the ratio
$c_\delta/v \propto G_\mu^{1/2}$ that is factored out of the
amplitude in eq.~(\ref{eq:LET}):
\beq
\label{eq:Kr}
K_r ~=~ \frac12 \left(\delta Z_{hh} + \frac{\delta G_\mu}{G_\mu}\right)~,
\eeq
where $\delta G_\mu$ is given in
eq.~(\ref{eq:dGmu}), and
\beq    
\delta Z_{hh} ~=~ 
\left.\frac{d\, \Pi_{hh}(p^2)}{dp^2}\right|_{p^2=0}~,
\eeq
$\Pi_{hh}(p^2)$ being the diagonal self-energy for the SM-like Higgs
scalar at external momentum $p^2$. An explicit formula for $\delta
Z_{hh}$ is given in the appendix~A.

\bigskip

Combining the four contributions in eq.~(\ref{eq:splitP2l}) we obtain
an explicit formula for ${\cal P}_h^{2\ell}$ that, by making use of
the tree-level expressions for the BSM Higgs masses in
eq.~(\ref{eq:massesnomix}), we can write in terms of $\lambda_{HT}$,
$\delta$, and the ratio $m_H^2/\mHpq$ (or, in alternative,
$\lambda_{HT}$, $\lambda_{T}$, and $\delta$). As the dependence on
$\delta$ is quite involved, we relegate this result to the
appendix~B. We remark here that ${\cal P}_h^{2\ell}$ does not depend
explicitly on the renormalization scale, which is consistent with the
fact that the parameter $c_\delta$ entering ${\cal P}_h^{1\ell}$ is
itself scale-independent when only the corrections controlled by
$\lambda_{HT}$ and $\lambda_{T}$ are considered. We also remark that
in the limit $\delta\rightarrow 0$ the BSM Higgs contribution to the
two-loop $h\gamma\gamma$ amplitude reduces to
\beq
\left.{\cal P}_h^{2\ell}\right|_{\delta\rightarrow 0}~=~
\frac{17\,\lambda_{HT}}{144\,\pi^2}~.
\label{eq:P2lsimpl}
\eeq
In view of the smallness of $v_T$ (and hence $\delta$) in realistic
EWSB scenarios, the approximation in eq.~(\ref{eq:P2lsimpl}) can be
expected to hold reasonably well even when $\delta\neq0$.  Remarkably,
eq.~(\ref{eq:P2lsimpl}) shows that, in the scenario~A, ${\cal
  P}_h^{2\ell}$ can become comparable in size with the charged-Higgs
contribution to ${\cal P}_h^{1\ell}$ when the coupling $\lambda_{HT}$
approaches the upper bounds discussed at the end of
section~\ref{sec:tree}.

\subsection{Calculation of the $h\gamma\gamma$ amplitude in the
  scenario B}
\label{sec:scenB}

In this scenario, characterized by the condition $\kappa_{HT}=v_T=0$,
the doublet and triplet scalars do not mix, the doublet has SM-like
couplings to fermions and to gauge bosons, and the charged and neutral
components of the triplet have degenerate masses at the tree level,
$\mHpq=\mHq=m_T^2 \,+\,\lambda_{HT}\,v^2/2$. The one-loop
$h\gamma\gamma$ amplitude in the limit of vanishing $m_h$ can again be
obtained from the combination of eqs.~(\ref{eq:LET}) and
(\ref{eq:Pigaga1l}), now setting $c_\delta=1$ in the
former. Differently from the case of the scenario~A, the amplitude
${\cal P}_h^{1\ell}$ can be truly decomposed into SM and BSM parts:
\bea
\label{eq:P1lSM}
{\cal P}_h^{1\ell,\,\smallSM} &=&\frac43 \,Q_t^2\,N_c ~-~ 7
~=~ -\frac{47}9~,\\[2mm]
\label{eq:P1lBSM1}
{\cal P}_h^{1\ell,\,\smallBSM} &=&
\frac{\lambda_{HT}\,v^2}{6\,\mHpq}\\[2mm]
&=& 
\frac{1}{3}\,\left(1-\frac{m_T^2}{\mHpq}\right)~.
\label{eq:P1lBSM2}
\eea
Beyond the LO, the two expressions for ${\cal
  P}_h^{1\ell,\,\smallBSM}$ in eqs.~(\ref{eq:P1lBSM1}) and
(\ref{eq:P1lBSM2}) are equivalent to each other when all of the
parameters involved are interpreted as $\msbar$, but we will see below
that the equivalence is broken for a different choice of
renormalization scheme.

\bigskip

Our results for the two-loop part of the $h\gamma\gamma$ amplitude are
remarkably compact in this scenario. For the BSM contribution to the
two-loop self-energy of the photon we find
\beq
\Pi_{\gamma\gamma}^{2\ell,\,\smallBSM}(0)~=~
-\frac{\alpha_{\rm em}}{192\pi^3}\,\left[\,\frac{\lambda^2_{HT}\,v^2}{2\,\mHpq}\,
  \left(1-2\,\ln\frac{\mHpq}{Q^2}\right)
  \,+\,5\,\lambda_T\,\left(1-\ln\frac{\mHpq}{Q^2}\right)\,\right]~,
\label{eq:Pigaga2l}
\eeq
where we denoted both the charged and neutral triplet masses by
$\mHpq$. Combining directly eq.~(\ref{eq:LET}) and
eq.~(\ref{eq:Pigaga2l}) we obtain for the genuine two-loop part of the
amplitude
\beq
    {\cal P}_h^{2\ell,\,\smallIPI}~=~
    \frac{\lambda_{HT}\,v^2}{96\pi^2\,\mHpq}\,\left\{    
    \lambda_{HT}\,\left[2-\frac{m_T^2}{\mHpq}\,
      \left(3-2\,\ln\frac{\mHpq}{Q^2}\right)\,\right]
    \,+\, 5\,\lambda_T \,\right\}~.
\label{eq:P2l-1PI}    
\eeq

\bigskip

We remark that in the scenario~B we do not need to introduce a
contribution to the two-loop $h\gamma\gamma$ amplitude analogous to
the one in eq.~(\ref{eq:shift}). Differently from the case of the
scenario~A, we do not use the minimum conditions of the scalar
potential to remove $m_T^2$ from the expression of the
$\msbar$-renormalized mass $\mHpq$ entering the one-loop amplitude,
and the condition of vanishing doublet-triplet mixing is preserved
beyond the tree level by the $Z_2$ symmetry.

The third contribution to the two-loop $h\gamma\gamma$ amplitude in
eq.~(\ref{eq:splitP2l}) accounts for our choices of renormalization
scheme for the parameters entering ${\cal P}_h^{1\ell,\,\smallBSM}$ in
eqs.~(\ref{eq:P1lBSM1}) and (\ref{eq:P1lBSM2}). We interpret the mass
of the charged Higgs boson as the pole mass $M^2_{H^\pm}$, which is related
to the $\msbar$ mass $\mHpq$ by
\beq
M^2_{H^\pm} ~=~ \mHpq \,+\, \Pi_{H^+H^-}(\mHpq)~,
\label{eq:MHpole}
\eeq
where the self-energy of the charged Higgs boson reads
\beq
\Pi_{H^+H^-}(\mHpq) ~=~ \frac{1}{16\pi^2}\,
\left[\,\lambda_{HT}^2\,v^2\,\left(\ln\frac{\mHpq}{Q^2}-2\right)
  \,+\,5\,\lambda_T\,\mHpq\,\left(\ln\frac{\mHpq}{Q^2}-1\right)\right]~.
\label{eq:PiHpHm}
\eeq
Concerning the remaining parameters in eqs.~(\ref{eq:P1lBSM1}) and
(\ref{eq:P1lBSM2}), we interpret $\lambda_{HT}$ and $m_T^2$ as
$\msbar$-renormalized parameters\,\footnote{An alternative scheme in
  which a parameter analogous to $m_T^2$ is connected to the
  renormalized $hHH$ vertex was considered in ref.~\cite{Aiko:2023nqj}
  for a variant of the THDM.}  at the generic renormalization scale
$Q$. The squared vev $v^2$ can be directly defined as
$(\sq2\,G_\mu)^{-1}$ without affecting the calculation of $\delta
{\cal P}_h^{1\ell}$, because in the scenario~B we find
$\Pi_{WW}^\smallBSM(0)=0$ under the approximations relevant to our
two-loop calculation, hence $\delta G_\mu=0$, see
eq.~(\ref{eq:dGmu}). However, the result for $\delta {\cal
  P}_h^{1\ell}$ depends on whether we choose to express the BSM
contribution to the one-loop amplitude as in eq.~(\ref{eq:P1lBSM1}) or
as in eq.~(\ref{eq:P1lBSM2}):
\bea
\left. \delta {\cal P}_h^{1\ell}\right|_{\lambda_{HT}} &=& 
\frac{\lambda_{HT}\,v^2}{6\,m_{H^\pm}^4}~\Pi_{H^+H^-}(\mHpq)~,\\[2mm]
\left. \delta {\cal P}_h^{1\ell}\right|_{m_T^2} &=& 
-\frac{m_T^2}{3\,m_{H^\pm}^4}~\Pi_{H^+H^-}(\mHpq)~.
\eea

\bigskip

In the scenario~B the parameter $K_r$ defined in eq.~(\ref{eq:Kr})
receives only the contribution of the WFR of the external Higgs field,
because $\delta G_\mu=0$ as mentioned above. We thus find
\beq
K_r ~=~ -\frac{\lambda_{HT}^2\,v^2}{128\pi^2\,\mHpq}~,
\label{eq:KrZ2}
\eeq
and the fourth contribution to the two-loop amplitude ${\cal
  P}_h^{2\ell}$ in eq.~(\ref{eq:splitP2l}) becomes $K_r\,({\cal
  P}_h^{1\ell,\,\smallSM} + {\cal P}_h^{1\ell,\,\smallBSM})$, where
the SM and BSM contributions to the one-loop amplitude are given in
eqs.~(\ref{eq:P1lSM})--(\ref{eq:P1lBSM2}).\footnote{Since this is
already a two-loop contribution, it does not matter whether we use
eq.~(\ref{eq:P1lBSM1}) or eq.~(\ref{eq:P1lBSM2}) for ${\cal
  P}_h^{1\ell,\,\smallBSM}$.}

\bigskip

Combining all of the results in
eqs.~(\ref{eq:P1lSM})--(\ref{eq:KrZ2}), we obtain two alternative
expressions for the total BSM contribution to the $h\gamma\gamma$
amplitude, i.e., ${\cal P}_h^{\smallBSM} = {\cal
  P}_h^{1\ell,\,\smallBSM} + {\cal P}_h^{2\ell}$, depending of whether
we express the one-loop part in terms of $\lambda_{HT}$ or of $m_T^2$:

\bea
    {\cal P}_h^{\smallBSM}\biggr|_{\lambda_{HT}} &=&
    \frac{\lambda_{HT}\,v^2}{6\,M^2_{H^\pm}}
    ~+~ \frac{\lambda_{HT}\,v^2}{96\pi^2\,\mHpq}\,\left[    
      \lambda_{HT}\,\left(\frac{35}{12}+2\,\ln\frac{\mHpq}{Q^2}\right)
      \,-\, \frac{5\lambda_{HT}^2\,v^2}{8\,\mHpq}      
      \,+\, 5\,\lambda_T \,\ln\frac{\mHpq}{Q^2}\,\right],\nn \\
    \label{eq:PBSMLHT}\\[3mm]
    {\cal P}_h^{\smallBSM}\biggr|_{m_T^2} &=&
    \frac{1}{3}\,\left(1-\frac{m_T^2}{M^2_{H^\pm}}\right)
    \,+~\frac{1}{48\pi^2}\,\left[\,
      \frac{(\mHpq-m_T^2)^2}{6\,\mHpq\,v^2}
      \,\left(68+15\,\frac{m_T^2}{\mHpq}\right)\right.\nn\\[1mm]
      &&~~~~~~~~~~~~~~~~~~~~~~~~~~~~~~~~~~~~~~~~~
      \left. +\,5\,\lambda_T\,
      \left(1-\frac{m_T^2}{\mHpq}\,\ln\frac{\mHpq}{Q^2}\right)\,\right].
    \label{eq:PBSMmT}
\eea    

\noindent
We remark that in the two-loop parts of eqs.~(\ref{eq:PBSMLHT}) and
(\ref{eq:PBSMmT}) we made use of the tree-level formula for the mass
of the charged Higgs boson to express either $m_T^2$ or $\lambda_{HT}$
in terms of the remaining parameters. Also, while the one-loop parts
of eqs.~(\ref{eq:PBSMLHT}) and (\ref{eq:PBSMmT}) are explicitly
expressed in terms of the pole mass $M^2_{H^\pm}$, in the two-loop
parts we denote the charged-Higgs mass as $\mHpq$, to stress that its
definition there amounts to a higher-order effect.

\bigskip

As we should expect from the overall scale independence of
$\Gamma(h\rightarrow\gamma\gamma)$, we find that the explicit scale
dependence of the two-loop part of ${\cal P}_h^{\smallBSM}$ in
eqs.~(\ref{eq:PBSMLHT}) and (\ref{eq:PBSMmT}) is compensated for by
the implicit scale dependence of the parameters entering the one-loop
part. The relevant terms in the RGEs of $\lambda_{HT}$ and $m_T^2$
are~\cite{Forshaw:2003kh}
\beq
\label{eq:RGEs}
\frac{d \lambda_{HT}}{d\ln Q^2} ~\supset~ \frac{\lambda_{HT}}{16\pi^2}\,
\left(2\,\lambda_{HT}\,+\,5\,\lambda_T\right)~,~~~~~~~~~~
\frac{d m_T^2}{d\ln Q^2} ~\supset~ \frac{5\,\lambda_{T}}{16\pi^2}\, m_T^2~,
\eeq
while the one-loop RGE for $v$ does not depend on the quartic Higgs
couplings, and the pole mass $M^2_{H^\pm}$ is obviously
scale-independent.

\bigskip

Finally, it might be worth discussing the behavior of ${\cal
  P}_h^{\smallBSM}$ in the two limiting cases $m_T^2\rightarrow 0$,
where $\mHpq\rightarrow \lambda_{HT}\,v^2/2$, and $m_T^2\rightarrow
\infty$, where also $\mHpq \rightarrow \infty$. For vanishing $m_T^2$,
the expression for ${\cal P}_h^{\smallBSM}$ that depends explicitly on
that parameter, eq.~(\ref{eq:PBSMmT}), becomes
\beq
{\cal P}_h^{\smallBSM}\biggr|_{m_T^2\rightarrow 0}
~=~ \frac13 ~+~
\frac{17\,\lambda_{HT}\,+\,15\,\lambda_{T}}{144\,\pi^2}~.
\label{eq:PBSMmT0}
\eeq
We note a discrepancy with the two-loop contribution to the
$h\gamma\gamma$ amplitude obtained in the scenario~A for the limit
$\delta\rightarrow 0$, see eq.~(\ref{eq:P2lsimpl}), in seeming
contradiction with the tree-level expectation that this limit should
correspond to $m_T^2\rightarrow0$. This is due to the fact that, at
the perturbative order required for consistency with the two-loop
calculation of the amplitude, the limit $\delta\rightarrow 0$ in the
scenario~A corresponds in fact to $m_T^2\rightarrow
5\,\lambda_{HT}\,\lambda_T\,v^2/(32\pi^2)$, as can be checked by
combining eqs.~(\ref{eq:min2}), (\ref{eq:renorm}), (\ref{eq:TH}) and
(\ref{eq:PihH}). If we take the latter limit in eq.~(\ref{eq:PBSMmT}),
we do indeed reproduce eq.~(\ref{eq:P2lsimpl}).
We also note that in the expression for ${\cal P}_h^{\smallBSM}$ that
depends on $\lambda_{HT}$, eq.~(\ref{eq:PBSMLHT}), the two-loop part
does not coincide with eq.~(\ref{eq:PBSMmT0}) when we set
$\mHpq=\lambda_{HT}\,v^2/2$. This is however compensated for by the
fact that the one-loop part does not in turn tend to $1/3$ for
$m_T^2\rightarrow 0$, because in that limit the pole mass
$M_{H^\pm}^2$ differs from $\lambda_{HT}\,v^2/2$ by a one-loop shift,
see eqs.~(\ref{eq:MHpole}) and (\ref{eq:PiHpHm}). In other words, the
expected limit is reproduced also in eq.~(\ref{eq:PBSMLHT}), but only
by the sum of the one- and two-loop parts.

In the opposite limit, $m_T^2\rightarrow \infty$, the expression for
${\cal P}_h^{\smallBSM}$ that depends on $\lambda_{HT}$ tends
manifestly to zero, reflecting the expected decoupling of the heavy
triplet states from the properties of the SM-like Higgs boson, because
all terms in eq.~(\ref{eq:PBSMLHT}) are suppressed by powers of
$v^2/\mHpq$. In contrast, the decoupling is not manifest in the
expression for ${\cal P}_h^{\smallBSM}$ that depends on $m_T^2$,
eq.~(\ref{eq:PBSMmT}), where the term proportional to $\lambda_T$ in
the two-loop part does not vanish. However, this is compensated for by
the one-loop part, which does not vanish either when $m_T^2\rightarrow
\infty$ due to a non-decoupling term in the relation between
$M^2_{H^\pm}$ and $\mHpq$, see eqs.~(\ref{eq:MHpole}) and
(\ref{eq:PiHpHm}). Once again, eq.~(\ref{eq:PBSMmT}) does reproduce
the expected decoupling of the heavy triplet states, but only in the
sum of the one- and two-loop parts.

An alternative way to make the decoupling of heavy states manifest in
the two-loop corrections even when using an expression for ${\cal
  P}_h^{\smallBSM}$ similar to eq.~(\ref{eq:PBSMmT}) was discussed in
refs.~\cite{Braathen:2019pxr, Braathen:2019zoh}, in the context of the
two-loop calculation of the Higgs self-couplings in the THDM. In those
studies it was proposed that the non-decoupling terms be absorbed in a
redefinition of the parameter that controls the mass of the heavy
states. Following that approach, we can define $(m_T^2)^{\rm dec}
\,=\, (m_T^2)^{\msbar} \,+\, \delta m_T^2$\,, where
\beq
\delta m_T^2 ~=~-\frac{5\,\lambda_T}{16\pi^2}\,m_T^2
\,\left(1-\ln\frac{m_T^2}{Q^2}\right)~,
\label{eq:mTdec}
\eeq
and we obtain for the BSM contribution to the $h\gamma\gamma$ amplitude
\bea
    {\cal P}_h^{\smallBSM}\biggr|_{(m_T^2)^{\rm dec}} &=&
    \frac{1}{3}\,\left(1-\frac{(m_T^2)^{\rm dec}}{M^2_{H^\pm}}\right)
    \,+~\frac{1}{48\pi^2}\,\left\{\,
      \frac{(\mHpq-m_T^2)^2}{6\,\mHpq\,v^2}
      \,\left(68+15\,\frac{m_T^2}{\mHpq}\right)\right.\nn\\[1mm]
      &&~~~~~~~~~~~~~~~~~~~~~~~~~~~~~~~~~~~~~~~~~
      \left. +\,5\,\lambda_T\,
      \left[1-\frac{m_T^2}{\mHpq}\,\left(1+\ln\frac{\mHpq}{m_T^2}
        \right)\,\right]\,\right\}~.
    \label{eq:PBSMmTdec}
\eea    
The two-loop part of ${\cal P}_h^{\smallBSM}$ in
eq.~(\ref{eq:PBSMmTdec}) now vanishes when $m_T^2\rightarrow\infty$\,,
because in that limit $\mHpq \approx m_T^2$ (again, we denote the mass
parameters in the two-loop part as $\mHpq$ and $m_T^2$ because their
definitions there amount to higher-order effects). On the other hand,
the vanishing of the one-loop part for $(m_T^2)^{\rm
  dec}\rightarrow\infty$ results from the combination of
eqs.~(\ref{eq:MHpole}), (\ref{eq:PiHpHm}) and (\ref{eq:mTdec}).

\section{Numerical impact of the two-loop BSM contributions to
  $\Gamma[h\rightarrow \gamma \gamma]$}
\vspace*{3mm}

\label{sec:numerics}

We now illustrate the numerical impact of the newly-computed two-loop
corrections on the prediction for $\Gamma[h\rightarrow \gamma \gamma]$
in the \TSM. We focus on the two scenarios described in the previous
sections, in which the neutral components of doublet and triplet do
not mix, so that the lighter neutral scalar $h$ has SM-like couplings
to fermions and gauge bosons and plays the role of the $125$-GeV Higgs
boson discovered at the LHC. A comprehensive analysis of the parameter
space of the model along the lines of
refs.~\cite{FileviezPerez:2008bj, Wang:2013jba, Chabab:2018ert,
  Bell:2020gug, Chiang:2020rcv, Ashanujjaman:2023etj,
  Butterworth:2023rnw, Ashanujjaman:2024pky,
  Crivellin:2024uhc}, taking into account all of the theoretical and
experimental constraints, is well beyond the scope of this paper. What
we aim to discuss here, at the qualitative level, is how the inclusion
of the two-loop corrections controlled by the couplings $\lambda_{HT}$
and $\lambda_T$ can affect the prediction for $\Gamma[h\rightarrow
  \gamma \gamma]$ in scenarios where those couplings are large. To
this purpose, we can neglect all of the two-loop corrections
controlled by the SM couplings, which are of ${\cal O}(1)$ or smaller.
The inclusion of those corrections as computed for the
SM~\footnote{See refs.~\cite{Zheng:1990qa, Djouadi:1990aj,
  Dawson:1992cy, Melnikov:1993tj, Djouadi:1993ji, Inoue:1994jq,
  Fleischer:2004vb} for the QCD part, refs.~\cite{Liao:1996td,
  Djouadi:1997rj, Fugel:2004ug, Degrassi:2005mc} for the EW
corrections involving the top quark, and refs.~\cite{Aglietti:2004nj,
  Aglietti:2004ki, Passarino:2007fp, Actis:2008ts} for the remaining
EW corrections.}  would refine our prediction for $\Gamma[h\rightarrow
  \gamma \gamma]$, but would not alter our qualitative conclusions on
the relevance of the two-loop corrections in scenarios with large
scalar couplings.

\bigskip

The LHC average of the signal strength for the production of the
$125$-GeV Higgs boson followed by its decay to a pair of photons
is~\cite{ParticleDataGroup:2022pth}
\beq
\label{eq:mugagaexp}
\mu_{\gamma\gamma}^{\rm exp} ~\equiv~
\frac{\sigma^{\rm exp}(pp\rightarrow h)\,
  {\rm BR}^{\rm exp}(h\rightarrow\gamma\gamma)}
     {\sigma^{\smallSM}(pp\rightarrow h)\,
       {\rm BR}^{\smallSM}(h\rightarrow\gamma\gamma)}
     ~=~1.10 \,\pm\,0.07~,
\eeq  
where the numerator of $\mu_{\gamma\gamma}^{\rm exp}$ contains the
experimental measurement, the denominator contains the SM prediction,
and the branching ratio (BR) for the decay $h\longrightarrow \gamma
\gamma$ is defined as the ratio of the partial width
$\Gamma[h\rightarrow \gamma \gamma]$ over the total width
$\Gamma_h^{\rm tot}$. Eq.~(\ref{eq:mugagaexp}) shows a mild
($1.4\sigma$) excess in $\mu_{\gamma\gamma}^{\rm exp}$, driven by the
CMS measurement~\cite{CMS:2022dwd}. If taken seriously, this excess
would suggest some tension with any model that predicts a positive BSM
contribution to the $h\gamma\gamma$ amplitude, and thus a value of
$\Gamma[h\rightarrow \gamma \gamma]$ lower than the SM prediction
(note that our conventions are such that the $h\gamma\gamma$ amplitude
is negative in the SM). In the case of the TSM, the one-loop
contribution from the charged Higgs boson has the same sign as
$\lambda_{HT}$. In the scenario~A, $\lambda_{HT}$ must always be
positive to avoid a tachyonic mass for the charged Higgs boson, see
eq.~(\ref{eq:massesnomix}), whereas in the scenario~B it can take
either sign, but negative values are constrained by the requirement
that the scalar potential be bounded from below, $\lambda_{HT}>
-\sqrt{2\,\lambda_H\lambda_T}$\,.
%
%
Even if the accumulation of additional data from the current and
future runs of the LHC should bring the central value of the signal
strength closer to the SM prediction, $\mu_{\gamma\gamma}^{\rm
  exp}\approx 1$, the comparison with the prediction of the TSM will
still provide significant constraints on the parameter space of the
model, also in view of the expected reduction of the
uncertainties~\cite{Cepeda:2019klc}.

To highlight the effect of the newly-computed two-loop corrections
controlled by $\lambda_{HT}$ and $\lambda_T$, we define simplified
predictions at one and two loops for the signal-strength parameter in
the TSM
\beq
\label{eq:defmu}
\mu_{\gamma\gamma}^{1\ell}~\equiv~\frac{\left| ~~{\cal
    F}_h^{1\ell}~~\right|^2} {\left|{\cal
    F}_h^{1\ell,\,\smallSM}\right|^2}~,~~~~~~~~~~
\mu_{\gamma\gamma}^{2\ell}~\equiv~\frac{ \left| {{\cal F}_h^{1\ell}}
   \left(1+\frac{{\cal P}_h^{2\ell}}{{\cal P}_h^{1\ell}}\right)\right|^2}
   {\left|{\cal F}_h^{1\ell,\,\smallSM}\right|^2}~,
\eeq
where ${\cal F}_h^{1\ell}$ and ${\cal F}_h^{1\ell,\,\smallSM}$ denote
the one-loop, $m_h$-dependent $h\gamma\gamma$ amplitudes in the TSM
and in the SM, respectively, and are given in the appendix C, whereas
${\cal P}_h^{1\ell}$ and ${\cal P}_h^{2\ell}$ are the one- and-
two-loop amplitudes for vanishing $m_h$ in the TSM, as computed in the
previous section. The way we organized the contributions in
eq.~(\ref{eq:defmu}) allows us to retain the Higgs-mass dependence of
the signal-strength parameter at the LO,\footnote{This improves the
definition of the simplified signal-strength parameter introduced in
our ref.~\cite{Degrassi:2023eii} for the THDM.} while keeping into
account that the two-loop corrections are known only in the limit of
vanishing $m_h$, and should be compared with a one-loop result
computed in the same limit to avoid introducing a bias.
Note that an implicit assumption in eq.~(\ref{eq:defmu}) is that the
large quartic Higgs couplings do not significantly affect the ratio of
production cross section over total decay width of the SM-like Higgs
boson, i.e., $[\sigma(pp\rightarrow h)\,/\,\Gamma_h^{\rm
    tot}]^{\smallTSM} \approx [\sigma(pp\rightarrow
  h)\,/\,\Gamma_h^{\rm tot}]^{\smallSM}$. Indeed, the dominant NLO
contributions involving those couplings affect the main production and
decay channels only through the common multiplicative factor $K_r$,
see eq.~(\ref{eq:Kr}), which cancels in the ratio.

\bigskip

\begin{figure}[t]
\begin{center}
  \vspace*{-1cm}
  \includegraphics[width=11cm]{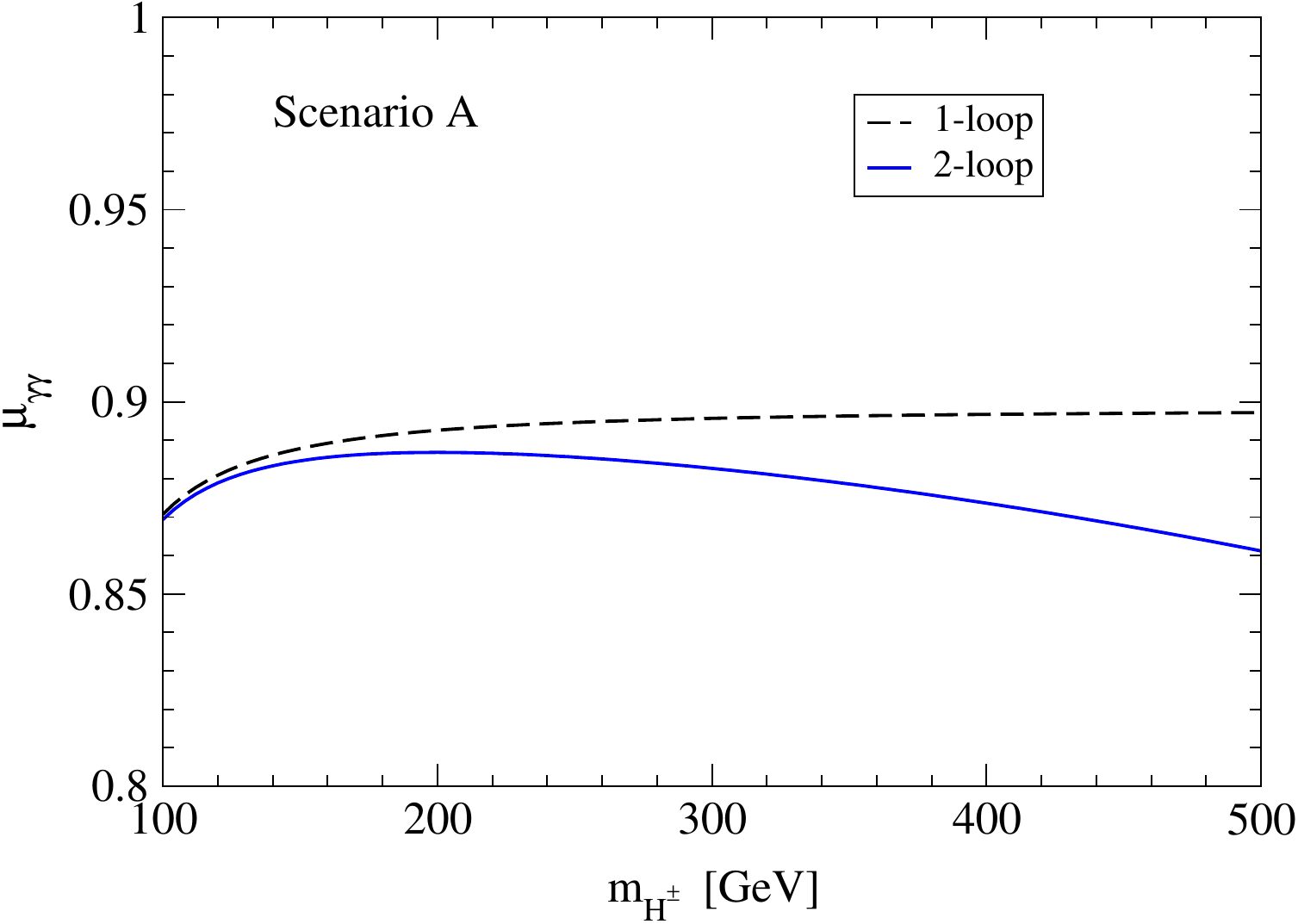}
  \caption{\em Signal strength $\mu_{\gamma\gamma}$ as a function of the
    charged-Higgs mass in the scenario~A of the TSM, with $v_T =
    6.5$~GeV. The meaning of the different lines is explained in the
    text.}
  \label{fig:scnA}
\end{center}
\end{figure}

In figure~\ref{fig:scnA} we plot the predictions for
$\mu_{\gamma\gamma}$ as a function of the charged-Higgs mass in the
scenario~A of the TSM. We denote the mass as $m_{H^\pm}$, because in
this scenario its definition beyond the tree level has only a
higher-order effect on our two-loop calculation of ${\cal
  P}_h^\smallBSM$.  We set $v_T =6.5$~GeV, close to the value that
leads to a prediction for $m_W$ in accordance with the CDF
measurement~\cite{CDF:2022hxs}. Even for this value of $v_T$, which
could be considered as an upper bound of the acceptable range,
eq.~(\ref{eq:cdGmu}) implies $c_\delta\approx 0.999$, and indeed we
find that our prediction for $\mu_{\gamma\gamma}^{2\ell}$, see
eq.~(\ref{eq:defmu}), would vary by less than $0.1\%$ if we used
$\left.{\cal P}_h^{2\ell}\right|_{\delta\rightarrow 0}$ from
eq.~(\ref{eq:P2lsimpl}) instead of the $\delta$-dependent result for
${\cal P}_h^{2\ell}$ given in the appendix~B. We can thus refer
directly to eq.~(\ref{eq:P2lsimpl}) when discussing the dependence of
the two-loop corrections on the quartic Higgs couplings. This implies
that, in the scenario A, the values of both $v_T$ and $\lambda_T$ have
a negligible impact on the prediction for the $h\gamma\gamma$
amplitude.

The black dashed line in figure~\ref{fig:scnA} represents the
prediction for $\mu_{\gamma\gamma}^{1\ell}$, whereas the blue solid
line represents the prediction for $\mu_{\gamma\gamma}^{2\ell}$.
Since in the scenario~A the couplings $\lambda_{HT}$ and $\lambda_T$
enter explicitly only the two-loop amplitude ${\cal P}_h^{2\ell}$, we
can extract $\lambda_{HT}$ from the tree-level relation $\mHpq =
\lambda_{HT}\,v^2/(2\,c_\delta^2)$, and we simply fix $\lambda_T=5$
(as mentioned above, the latter choice has a negligible impact on our
results). The figure shows that the one-loop prediction for the signal
strength reaches a plateau as soon as $m_{H^\pm}$ is large enough that
the BSM contribution to the $h\gamma\gamma$ amplitude is well
approximated by the result in the limit of vanishing $m_h$, i.e.,
${\cal P}_h^{1\ell\,,\smallBSM}= c_{\delta}/3$. The effect of the
two-loop BSM contributions manifests itself as a negative correction
to the signal strength that increases in size with growing
$m_{H^\pm}$. In view of the relation between $m_{H^\pm}$ and
$\lambda_{HT}$ in the scenario~A, this behavior can be directly traced
back to the dependence of the two-loop corrections on $\lambda_{HT}$
in the r.h.s.~of eq.~(\ref{eq:P2lsimpl}). We note that the largest
value of $m_{H^\pm}$ considered in figure~\ref{fig:scnA} corresponds
to $\lambda_{HT}\approx 8$, well within the upper bound from
perturbative unitarity.

Comparing directly the TSM predictions in figure~\ref{fig:scnA} with
the measured value for the signal strength in eq.~(\ref{eq:mugagaexp})
would lead to the rather drastic conclusion that the scenario~A is
ruled out at the $3\sigma$ level for all values of $m_{H^\pm}$ when
the two-loop contributions are included (in contrast, the one-loop
prediction differs from the measured value by slightly less than
$3\sigma$ as soon as $m_{H^\pm}\gtrsim 170$~GeV). If we assume instead
that the excess in $\mu_{\gamma\gamma}^{\rm exp}$ is due to a
statistical fluctuation, and that more data will not only shrink the
experimental uncertainty band but also bring the central value of the
signal strength close to one, a future $2\sigma$ exclusion line could
fall right into the region spanned by the two lines in
figure~\ref{fig:scnA}. The two-loop corrections might thus turn a
scenario that was marginally allowed into an excluded one. In any
case, it is clear that for large values of $\lambda_{HT}$ the
inclusion of the two-loop corrections is necessary to obtain an
accurate prediction for the $h\longrightarrow\gamma\gamma$ signal
strength.


\begin{figure}[t]
\begin{center}
  \vspace*{-1cm}
  \includegraphics[width=8.3cm]{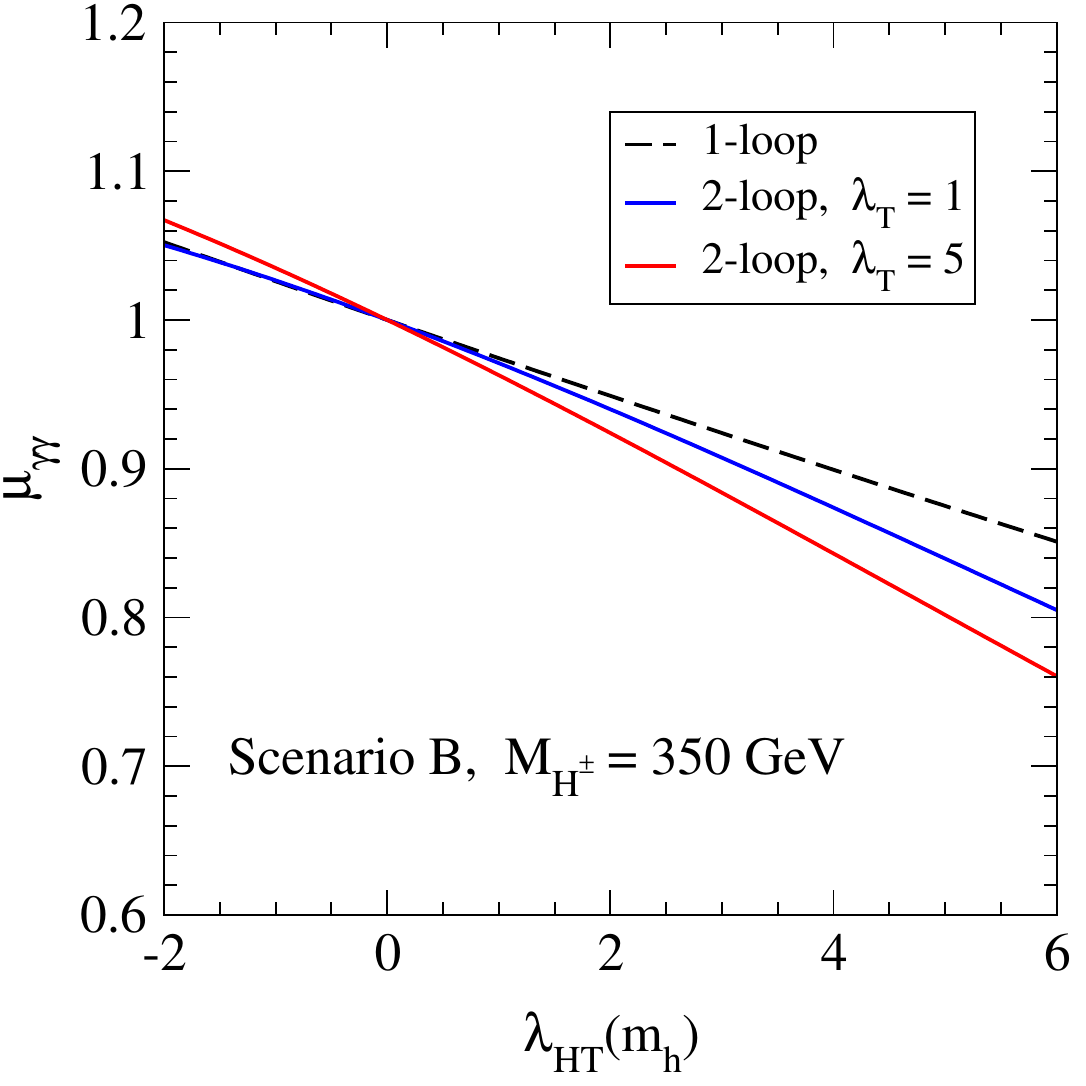}~~~~
  \includegraphics[width=8.3cm]{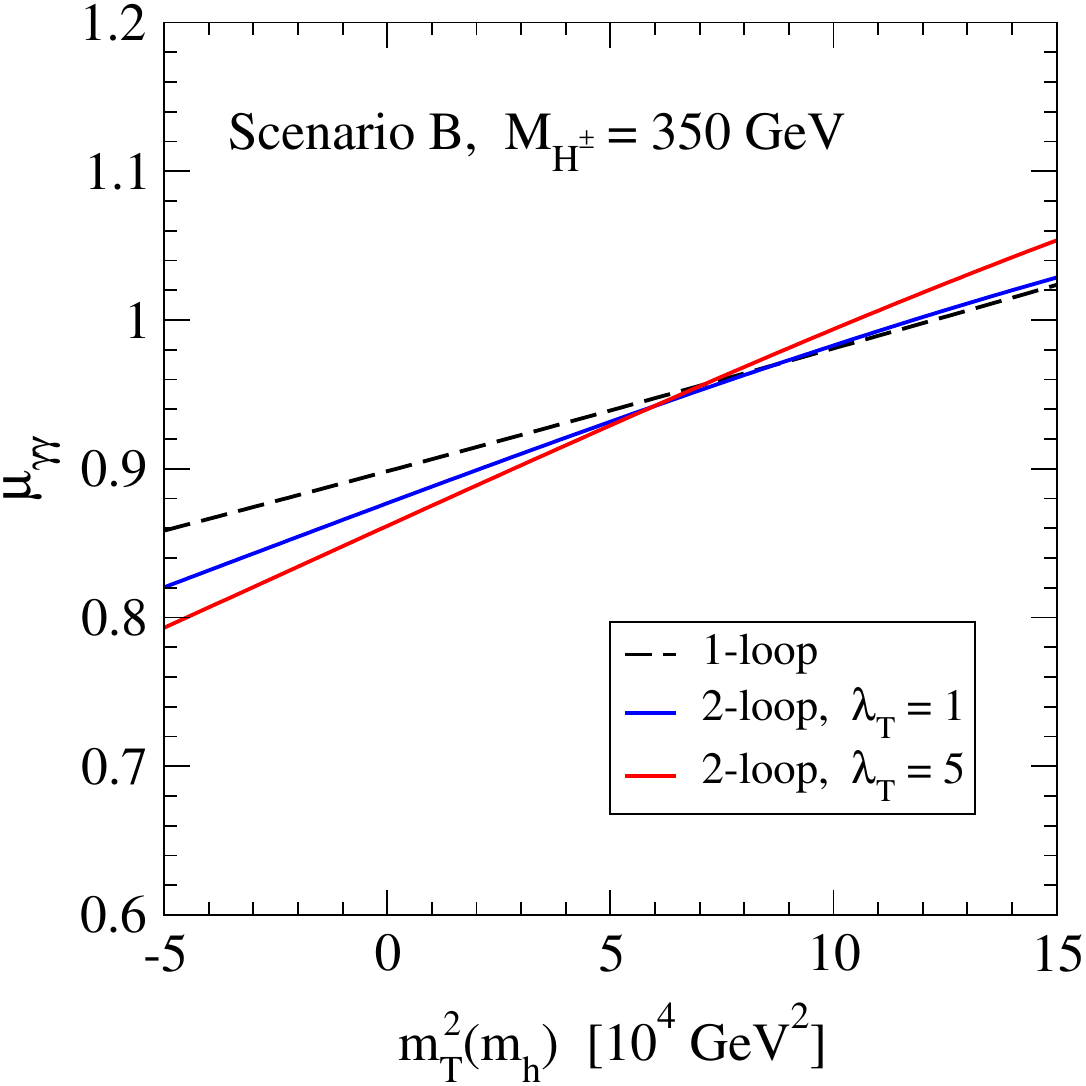}
  \caption{\em Signal strength $\mu_{\gamma\gamma}$ as a function of
    $\lambda_{HT}(m_h)$ (left) or $m_T^2(m_h)$ (right) in the
    scenario~B of the TSM, with $M_{H^\pm}=350$~GeV. The meaning of
    the different lines is explained in the text.}
  \label{fig:scenB-Qh}
  \vspace*{-11mm}
\end{center}
\end{figure}

\hfill

Moving on to the scenario~B, in figure~\ref{fig:scenB-Qh} we plot the
predictions for $\mu_{\gamma\gamma}$ as a function of either
$\lambda_{HT}$ (left plot) or $m_T^2$ (right plot), both interpreted
as $\msbar$-renormalized parameters at the scale $Q=m_h$. We fix the
pole mass of the charged Higgs boson, which enters the one-loop part
of the $h\gamma\gamma$ amplitude in eqs.~(\ref{eq:PBSMLHT}) and
(\ref{eq:PBSMmT}), to $M_{H^\pm} = 350$~GeV, a value that according to
refs.~\cite{Bell:2020gug, Chiang:2020rcv} should be within reach of
the current searches for disappearing tracks at the LHC. In each plot,
the black dashed line represents the prediction for
$\mu_{\gamma\gamma}^{1\ell}$, whereas the blue and red solid lines
represent the predictions for $\mu_{\gamma\gamma}^{2\ell}$ when
$\lambda_T=1$ and $\lambda_T=5$, respectively. The requirements that
the scalar potential be bounded from below and that the SM-like
minimum be global imply $-0.7\lesssim \lambda_{HT}\lesssim 4.8$ and
$-2\!\times\! 10^4~{\rm GeV}^2 \lesssim m_T^2\lesssim 14\!\times\!
10^4~{\rm GeV}^2$ when $\lambda_T=1$, and $-1.6\lesssim
\lambda_{HT}\lesssim 5.7$ and $-5\!\times\! 10^4~{\rm GeV}^2 \lesssim
m_T^2\lesssim 17\!\times\! 10^4~{\rm GeV}^2$ when
$\lambda_T=5$. However, we stress that these bounds are derived from
tree-level relations and should be treated as merely qualitative, and
that the upper (lower) bounds on $\lambda_{HT}$ (on $m_T^2$) could be
somewhat relaxed by allowing for a metastable SM-like vacuum.

In the left plot in figure~\ref{fig:scenB-Qh}, the comparison between
the three lines shows that the two-loop corrections can be quite
relevant in this scenario, and that their size increases for
increasing $\lambda_T$. For example, if we assumed that the ``true''
central value of the signal strength is exactly one, the current
$2\sigma$ exclusion line would sit at $\mu_{\gamma\gamma}=0.86$. The
comparison with the TSM prediction would thus rule out
$\lambda_{HT}\gtrsim 5.6$ when based on the one-loop calculation of
the $h\gamma\gamma$ amplitude, $\lambda_{HT}\gtrsim 4.4$ when based on
the two-loop calculation with $\lambda_T=1$, and $\lambda_{HT}\gtrsim
3.6$ when based on the two-loop calculation with $\lambda_T=5$.
The impact of the two-loop corrections in the right plot in
figure~\ref{fig:scenB-Qh} is qualitatively similar to the one shown in
the left plot: a $2\sigma$ exclusion line at $\mu_{\gamma\gamma}=0.86$
would rule out $m_T^2\lesssim -5\!\times\!10^{-4}$~GeV$^2$ when based
on the one-loop calculation, $m_T^2\lesssim
-1.5\!\times\!10^{-4}$~GeV$^2$ when based on the two-loop calculation
with $\lambda_T=1$, and $m_T^2 \lesssim 0$ when based on the two-loop
calculation with $\lambda_T=5$.

\bigskip

\begin{figure}[t]
\begin{center}
  \vspace*{-1cm}
  \includegraphics[width=8.3cm]{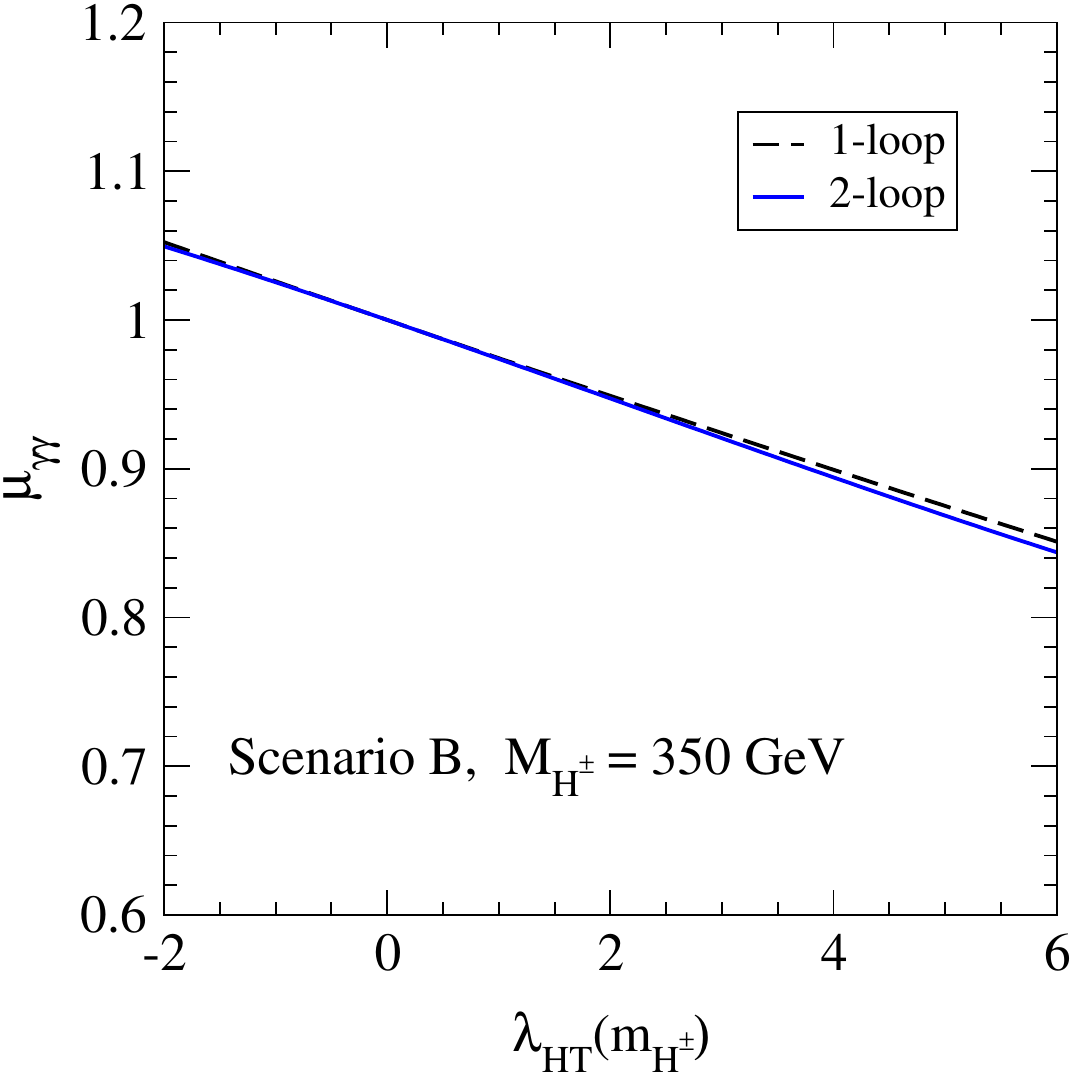}~~~~
  \includegraphics[width=8.3cm]{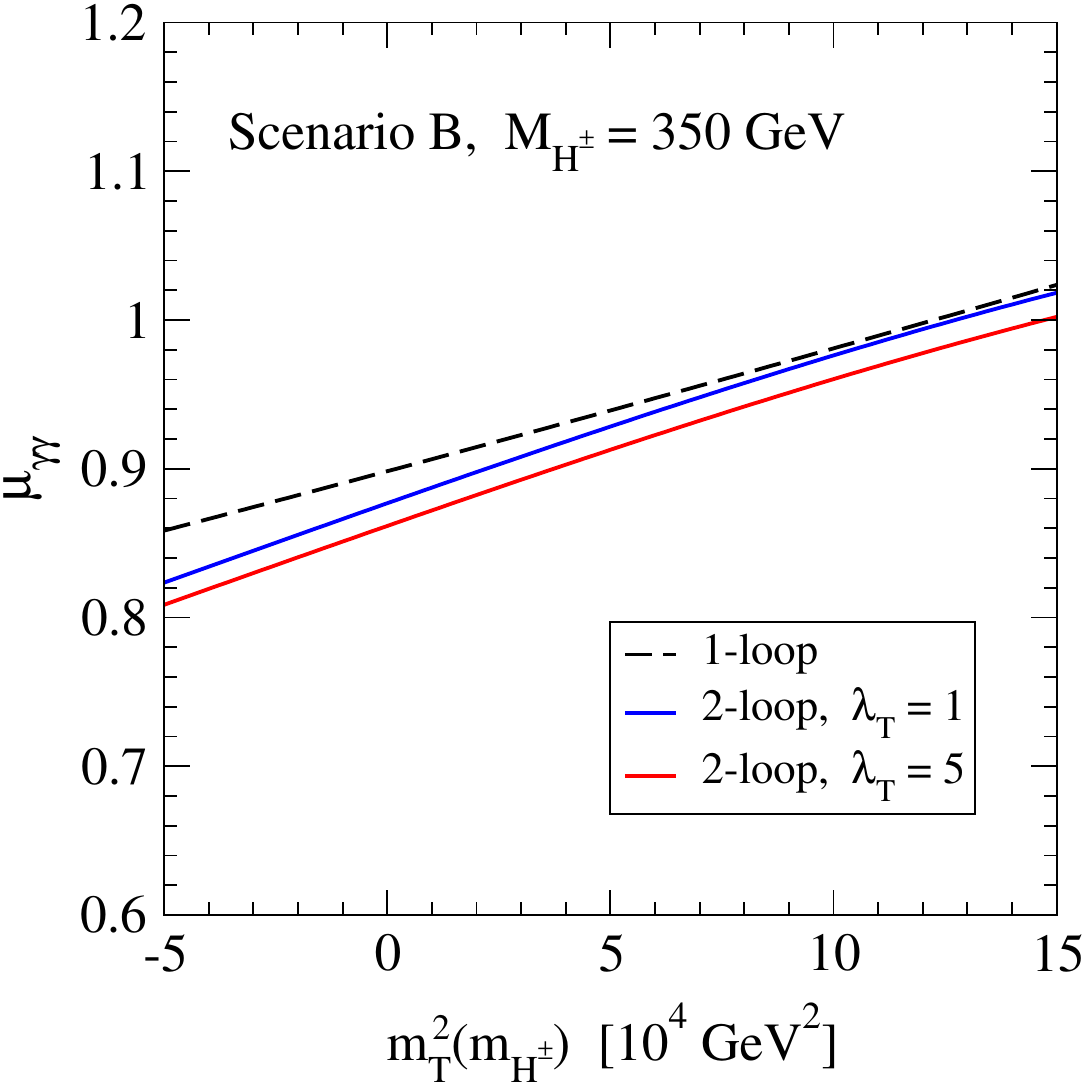}
  \caption{\em Same as figure~\ref{fig:scenB-Qh}, but the
    parameters on the $x$-axis are computed at $Q=m_{H^\pm}$.}
  \label{fig:scenB-QHp}
\vspace*{-5mm}
\end{center}
\end{figure}

Finally, in figure~\ref{fig:scenB-QHp} we plot the same quantities as
in figure~\ref{fig:scenB-Qh}, but as function of $\msbar$-renormalized
parameters at the scale $Q=m_{H^\pm}$ instead of $Q=m_{h}$ (note that
changing the input scale amounts to considering a different region of
the TSM parameter space). We see that in the left plot of
figure~\ref{fig:scenB-QHp} -- where the signal strength is shown as a
function of $\lambda_{HT}(m_{H^\pm})$ -- the two-loop corrections are
now significantly smaller, and they do not depend at all on
$\lambda_T$. Inspection of eq.~(\ref{eq:PBSMLHT}) shows indeed that
the $\lambda_T$-dependent part of ${\cal P}_h^{2\ell}$ is entirely
proportional to $\ln(\mHpq/Q^2)$, and that there happens to be a
significant cancellation between the non-logarithmic terms in the
$\lambda_T$-independent part. Therefore, taking
$\lambda_{HT}(m_{H^\pm})$ as input parameter in phenomenological
analyses of this scenario might look like the most convenient choice,
because it reduces the dimensionality of the parameter space on which
the two-loop prediction for $\mu_{\gamma\gamma}$ depends. In contrast,
we see that in the right plot -- where the signal strength is shown as
a function of $m_T^2(m_{H^\pm})$ -- the two-loop corrections remain
significant. Inspection of eq.~(\ref{eq:PBSMmT}) shows indeed that in
this case the choice $Q=m_{H^\pm}$ does not entirely kill off the
$\lambda_T$-dependent part of ${\cal P}_h^{2\ell}$, and that the
$\lambda_T$-independent part does not depend at all on the scale. This
said, we stress that the choice of input parameters always depends on
the kind of analysis that is being performed, and that there might be
compelling reasons to pick a parameter other than
$\lambda_{HT}(m_{H^\pm})$.

\vspace*{1mm}
\section{Conclusions}
\label{sec:conclusions}
\vspace*{1mm}

The requirement that an extension of the SM accommodate a scalar with
properties compatible with those observed at the LHC constrains the
parameter space of the BSM model even before the direct observation of
any new particles. In models with an extended Higgs sector, some of
the quartic scalar couplings can be significantly larger than any of
the SM couplings, while still satisfying all of the theoretical and
experimental constraints. In such cases, the radiative corrections
controlled by those couplings can be expected to be the dominant ones,
and their inclusion may prove necessary in order to obtain accurate
predictions for the properties of the observed Higgs boson.

In this paper we computed the two-loop BSM contributions to $\Gamma(
h\rightarrow \gamma \gamma)$ in the TSM, a model in which the Higgs
sector of the SM is augmented with a real triplet of $SU(2)$.  We
considered two scenarios in which the neutral components of doublet
and triplet do not mix, so that the lighter neutral scalar $h$ has (at
least approximately) SM-like couplings to fermions and gauge
bosons. Following the approach of our ref.~\cite{Degrassi:2023eii},
where an analogous calculation was performed for the THDM, we focused
on the corrections controlled by the potentially large scalar
couplings, and we made use of a LET that connects the $h \gamma
\gamma$ amplitude to the derivative of the photon self-energy
w.r.t.~the vev of the SM-like Higgs field. This allowed us to obtain
explicit and compact formulas for the two-loop BSM contributions to
the $h\gamma\gamma$ amplitude in the two considered scenarios.

\bigskip

After describing our calculation, we briefly illustrated the numerical
impact of the newly-computed two-loop BSM contributions. We chose not
to embark in an extensive analysis of the TSM parameter space, but
rather discuss, at the qualitative level, how the inclusion of the
two-loop corrections controlled by the quartic scalar couplings can be
necessary in order to obtain a precise prediction for
$\Gamma[h\rightarrow \gamma \gamma]$ in scenarios where those
couplings are large. We defined a simplified signal-strength parameter
$\mu_{\gamma\gamma}$, and showed how the inclusion of the two-loop BSM
contributions tends to exacerbate the tension between the measured
value of the signal strength, which is in fact slightly above the SM
prediction, and the prediction of the TSM in the considered scenarios,
which is somewhat below the SM prediction. We also discussed how, in
one of the considered scenarios, the relevance of the two-loop
corrections depends on which parameters are used as input for the
one-loop part of the $h\gamma\gamma$ amplitude, and we identified a
choice that might prove more convenient than the others.

What we presented here is, to the best of our knowledge, the first
calculation of two-loop corrections to the properties of the Higgs
boson in the TSM, and our investigation could obviously be extended in
multiple directions. To start with, our results for
$\Gamma[h\rightarrow \gamma \gamma]$ could be generalized to the case
of a non-zero (but necessarily small) mixing between the neutral
components of doublet and triplet, which would entail a number of
complications related to the renormalization of the mixing
angle. Beyond $\Gamma[h\rightarrow \gamma \gamma]$, the techniques we
developed could also be used to compute the two-loop BSM contributions
to the $\rho$ parameter, in analogy with the THDM calculation of
refs.~\cite{Hessenberger:2016atw, Hessenberger:2022tcx}, and to the
trilinear self-coupling $\lambda_{hhh}$, in analogy with the THDM
calculation of refs.~\cite{Braathen:2019pxr, Braathen:2019zoh}. We
hope that the results presented in this paper will help the collective
effort to use the properties of the Higgs boson as a probe of what
lies beyond the SM.

\vspace*{1mm}
\section*{Acknowledgments}

We thank M.~Goodsell for useful communications about the
implementation of the TSM in {\tt SARAH}. 


\section*{Appendix A: One-loop tadpoles and self-energies in the scenario A}
\setcounter{equation}{0}
\renewcommand{\theequation}{A\arabic{equation}}

In this appendix we list explicit formulas for the finite parts of the
one-loop tadpoles and self-energies that are relevant to our
calculation. They were obtained by adapting to the TSM the general
formulas given in refs.~\cite{Braathen:2016cqe, Braathen:2018htl},
under the limits of vanishing doublet-triplet mixing angle $\gamma$,
vanishing SM-like Higgs mass, and vanishing gauge and Yukawa couplings
(except for an overall factor $g^2$ in the BSM contribution to the $W$
self-energy).

\bea
16\pi^2\,T_h &=& -\frac{\lambda_{HT}\,v}{2}
\left[2\,\mHpq\,\left(1-\ln\frac{\mHpq}{Q^2}\right)
  \,+\,\mHq\,\left(1-\ln\frac{\mHq}{Q^2}\right)\right]~,\\[4mm]
\label{eq:TH}
16\pi^2\,T_H &=&
\left(\lambda_{HT}\,s_\delta^2+\lambda_T\,c_\delta^2\right)\,v\,t_\delta\,
\mHpq\,\left(1-\ln\frac{\mHpq}{Q^2}\right)\nn\\[1mm]
&& +~
\frac32\,\lambda_T\,v\,t_\delta\,\mHq\,\left(1-\ln\frac{\mHq}{Q^2}\right)~,
\\[4mm]
\label{eq:Pihh}
16\pi^2\,\Pi_{hh}(0) &=&
-\frac{\lambda_{HT}}{2}
\left[2\,c_\delta^2\,\mHpq\,\left(1-\ln\frac{\mHpq}{Q^2}\right)
  \,+\,\mHq\,\left(1-\ln\frac{\mHq}{Q^2}\right)\right]\nonumber\\[4mm]
&&+~ \frac{\lambda_{HT}^2\,v^2}{2}\left[2\,\ln\frac{\mHpq}{Q^2}
  \,+\,\ln\frac{\mHq}{Q^2} - t_\delta^2\left(1-\ln\frac{\mHpq}{Q^2}
  \right)\right]~,\\[4mm]
\label{eq:PihH}
16\pi^2\,\Pi_{hH}(0) &=&
-\frac{\lambda_{HT}\,v^2\,t_\delta}{2} \left[
  2\,(\lambda_{HT}-\lambda_T)\,s_\delta^2
  +\lambda_T\left(2\,\ln\frac{\mHpq}{Q^2} + 3\,\ln\frac{\mHq}{Q^2}\right)
  \right]\,,\\[4mm]
16\pi^2\,\left.\frac{d\Pi_{hh}(p^2)}{d p^2}\right|_{p^2=0} &=&
-\frac{\lambda_{HT}^2\,v^2}{12}\left[\frac{2 + 3\,t_\delta^2}{\mHpq} \,+\,
  \frac{1}{\mHq}\right]~,\\[4mm]
16\pi^2\,\Pi_{WW}^\smallBSM(0) &=&
-\frac{g^2}{8}\,\left[
  (3+c_{2\delta})\,\mHpq + 4\,\mHq +
  8\,c_\delta^2\,\frac{\mHpq\,\mHq}{\mHpq-\mHq}\,\ln\frac{\mHq}{\mHpq}\right]~.
\eea
Note that our convention for the sign of the scalar self-energies
implies $M_\phi^2 = (m_\phi^2)^{\msbar} + \Pi_{\phi\phi}(m^2_\phi)$,
whereas our convention for the sign of the W self-energy implies
$M_W^2 = (m_W^2)^{\msbar} - \Pi_{WW}(m^2_W)$.

\vfill
\newpage

\section*{Appendix B: Two-loop $h\gamma\gamma$ amplitude in the scenario A}
\setcounter{equation}{0}
\renewcommand{\theequation}{B\arabic{equation}}

We provide here an explicit formula for the two-loop $h\gamma\gamma$
amplitude ${\cal P}_h^{2\,\ell}$, valid under the assumptions of
vanishing mixing between the neutral components of doublet and
triplet, vanishing mass of the SM-like Higgs, and vanishing gauge and
Yukawa couplings. We made use of the tree-level expressions for the
BSM Higgs masses in eq.~(\ref{eq:massesnomix}) to write ${\cal
  P}_h^{2\ell}$ in terms of $\lambda_{HT}$, $\delta$\,, and the ratio
$x\equiv \mHq/\mHpq$\,. We remark that the latter can also be
expressed as $x = c_\delta^2 + s_\delta^2 \,\lambda_T/\lambda_{HT}$\,,
in which case $\lambda_T$ is treated as an input parameter instead of
$x$.

\bea
    {\cal P}_h^{2\,\ell} &=&
    \frac{\lambda_{HT}\,c_\delta\,s_\delta^{-2}}{864\,\pi^2\,x\,(x-4)^2}\,\biggr\{
    ~520-8736\,x+1611\,x^2+800\,x^3-162\,x^4+9\,(2\,x^2+3)(x-4)^2\,c_{2\delta}
    \nn\\[2mm]
    && ~~~~~~~~~~~~~~~~~~~~~~~~
    -\,2\,\left(560+1752\,x-612\,x^2-71\,x^3+18\,x^4\right)\,c_\delta^2\nn\\[2mm]
    && ~~~~~~~~~~~~~~~~~~~~~~~~
    +\,6\,\left(64+1800\,x-921\,x^2+128\,x^3\right)\,c_\delta^4\nn\\[2mm]
    && ~~~~~~~~~~~~~~~~~~~~~~~~
    +\,3\,x\,(1+2\,x)\,(x-4)^2\,(19+11\,c_{2\delta})\,c_{\delta}^{-4}
    \nn\\[2mm]
    && ~~~~~~~~~~~~~~~~~~~~~~~~
    -\frac{12\,\ln x}{(x-1)(x-4)}\,\biggr[
      ~36+192\,x+1703\,x^2-1401\,x^3+363\,x^4-29\,x^5
    \nn\\[2mm]
    && ~~~~~~~~~~~~~~~~~~~~~~~~~~~~~~~~~~~~~~~~~~~~~~~~~
    -6\left(24-548\,x+548\,x^2-228\,x^3+47\,x^4-5\,x^5\right)\,c_{\delta}^2
    \nn\\[2mm]
    && ~~~~~~~~~~~~~~~~~~~~~~~~~~~~~~~~~~~~~~~~~~~~~~~~~
    +12\left(12-308\,x+210\,x^2-57\,x^3+8\,x^4\right)\,c_{\delta}^4
    \nn\\[2mm]
    && ~~~~~~~~~~~~~~~~~~~~~~~~~~~~~~~~~~~~~~~~~~~~~~~~~
    +8\,x^2\,(x-4)^3\,c_{\delta}^{-2}\biggr]
    \nn\\[2mm]
    && ~~~~~~~~~~~~~~~~~~~~~~~~
    +\frac{216\,\phi(x/4)}{x\,(x-4)}\,\biggr[
     ~2+7\,x-8\,x^2+2\,x^3   
    \nn\\[2mm]
    && ~~~~~~~~~~~~~~~~~~~~~~~~~~~~~~~~~~~~~~~~~~~~~
     -2\,\left(4+8\,x-14\,x^2+6\,x^3-x^4\right)\,c_{\delta}^2
    \nn\\[2mm]
    && ~~~~~~~~~~~~~~~~~~~~~~~~~~~~~~~~~~~~~~~~~~~~~
    + \left(8+4\,x-14\,x^2+5\,x^3\right)\,c_{\delta}^4~
    \biggr]\,\biggr\}~.
\label{eq:P2l-A}    
\eea    
The function $\phi(z)$ entering eq.~(\ref{eq:P2l-A}) above is
defined as
\beq
\phi(z) = \left\{
\begin{tabular}{ll}
  $4\, \sqrt{\frac{z}{1-z}} ~{\rm Cl}_2 ( 2 \arcsin \sqrt z )$ \, , &  
  $(0 < z < 1)$ \, , \\ \\
  ${ \frac1{\lambda} \left[ - 4 \,{\rm Li_2} (\frac{1-\lambda}2) +
    2\, \ln^2 (\frac{1-\lambda}2) - \ln^2 (4z) +\pi^2/3 \right] }$ \, ,
  & ~~~$(z \ge 1)$ \,,
\end{tabular}
\label{eq:phi}
\right.
\eeq
where ${\rm Cl}_2(z)= {\rm Im} \,{\rm Li_2} (e^{iz})$ is the Clausen
function, and $\lambda = \sqrt{1 - (1 / z)}$.

\bigskip

\vfill
\newpage

\section*{Appendix C: Higgs-mass dependence of the one-loop $h\gamma\gamma$ amplitude}
\setcounter{equation}{0}
\renewcommand{\theequation}{A\arabic{equation}}

We provide here explicit formulas for the one-loop, Higgs-mass
dependent $h\gamma\gamma$ amplitudes entering eq.~(\ref{eq:defmu}),
for the TSM scenarios A and B as well as for the SM. They read
\bea
\label{eq:FhA}
\left.{\cal F}_h^{1\ell}\right|^{\rm A}_{\phantom {m_T^2}} &=&
    c_\delta\,F_0\left(\frac{4\,m_{H^\pm}^2}{m_h^2}\right)
    \,+\, Q_t^2\,N_c\,c_\delta^{-1}\,F_{1/2}\left(\frac{4\,m_t^2}{m_h^2}\right)
    \,+\, c_\delta\, F_1\left(\frac{4\,m_W^2}{m_h^2}\right)
    ~,\\[3mm]
\label{eq:FhB1}
\left.{\cal F}_h^{1\ell}\right|^{\rm B}_{\lambda_{HT}} \!\!&=&
    \frac{\lambda_{HT}\,v^2}{2\,M_{H^\pm}^2}\,
    F_0\left(\frac{4\,m_{H^\pm}^2}{m_h^2}\right)
    \,+\, Q_t^2\,N_c\,F_{1/2}\left(\frac{4\,m_t^2}{m_h^2}\right)
    \,+\, F_1\left(\frac{4\,m_W^2}{m_h^2}\right)
    ~,\\[3mm]
\label{eq:FhB2}
\left.{\cal F}_h^{1\ell}\right|^{\rm B}_{m_T^2} &=&
    \left(1-\frac{m_T^2}{M_{H^\pm}^2}\right)
    F_0\left(\frac{4\,m_{H^\pm}^2}{m_h^2}\right)
    \,+\, Q_t^2\,N_c\,F_{1/2}\left(\frac{4\,m_t^2}{m_h^2}\right)
    \,+\, F_1\left(\frac{4\,m_W^2}{m_h^2}\right)
    ~,\\[3mm]
\label{eq:FhSM}
    {\cal F}_h^{1\ell\,,\smallSM} &=&
    Q_t^2\,N_c\,F_{1/2}\left(\frac{4\,m_t^2}{m_h^2}\right)
    \,+\, F_1\left(\frac{4\,m_W^2}{m_h^2}\right)~,
\eea
where for the scenario~B we distinguish the two cases in which the
one-loop part is expressed in terms of either $\lambda_{HT}$ or
$m_T^2$.  The loop functions $F_0$, $F_{1/2}$ and $F_1$ are
\bea
\label{eq:F0}
F_0(x) &=& -x \left[1-x\left(\arcsin \frac 1{\sqrt x}\right)^2\,\right]~,\\[2mm]
F_{1/2}(x) &=& 2\,x\left[1+(1-x)\left(\arcsin\frac 1{\sqrt x} \right)^2\,
  \right]~,\\[2mm]
\label{eq:F1}
F_1(x) &=& -2 - 3\,x\left[1+(2-x)
  \left(\arcsin\frac 1{\sqrt x}\right)^2\,\right]~.
\eea
In eqs.~(\ref{eq:F0})--(\ref{eq:F1}) we provide only the
representations of the loop functions valid when $x\geq1$, because we
neglect the tiny contributions of the light fermions and we consider
only scenarios where $m_{H^\pm}> m_h/2$. The limits of the loop
functions for $x\rightarrow \infty$ are $F_0(x)\rightarrow 1/3$,
$F_{1/2}(x)\rightarrow 4/3$, and $F_1(x)\rightarrow -7$, so that for
vanishing $m_h$ we recover eqs.~(\ref{eq:P1l-A}) and (\ref{eq:P1lBSM})
for the scenario~A and eqs.~(\ref{eq:P1lSM})--(\ref{eq:P1lBSM2}) for
the scenario~B.  Our notation for the mass of the charged Higgs boson
is meant to stress that its one-loop definition is relevant only for
the pre-factor of $F_0$ in eqs.~(\ref{eq:FhB1}) and (\ref{eq:FhB2}),
where it must be interpreted as the pole mass. For the SM-particle
masses entering eqs.~(\ref{eq:FhA})--(\ref{eq:FhSM}) and for the Fermi
constant\,\footnote{The value of $G_\mu$ is needed to obtain
$c_\delta$ in the scenario A and $v$ in the scenario B.} we use the
current PDG values~\cite{ParticleDataGroup:2022pth}, i.e., $m_h =
125.25$~GeV, $m_t = 172.69$~GeV, $m_W=80.377$~GeV, and
$G_\mu=1.16638\!\times\!10^{-5}$~GeV$^{-2}$.

\vfill
\newpage

\bibliographystyle{utphys}
\bibliography{triplet}

\providecommand{\href}[2]{#2}\begingroup\raggedright\begin{thebibliography}{10}

\bibitem{CMS:2012qbp}
{\bf CMS} Collaboration, S.~Chatrchyan {\em et al.}, {\em {Observation of a New
  Boson at a Mass of 125 GeV with the CMS Experiment at the LHC}}.
  \href{http://dx.doi.org/10.1016/j.physletb.2012.08.021}{Phys. Lett. B {\bf
  716} (2012)  30--61}, \href{http://arxiv.org/abs/1207.7235}{{\tt
  arXiv:1207.7235 [hep-ex]}}.

\bibitem{ATLAS:2012yve}
{\bf ATLAS} Collaboration, G.~Aad {\em et al.}, {\em {Observation of a new
  particle in the search for the Standard Model Higgs boson with the ATLAS
  detector at the LHC}}.
  \href{http://dx.doi.org/10.1016/j.physletb.2012.08.020}{Phys. Lett. B {\bf
  716} (2012)  1--29}, \href{http://arxiv.org/abs/1207.7214}{{\tt
  arXiv:1207.7214 [hep-ex]}}.

\bibitem{ParticleDataGroup:2022pth}
{\bf Particle Data Group} Collaboration, R.~L. Workman {\em et al.}, {\em
  {Review of Particle Physics}}.
  \href{http://dx.doi.org/10.1093/ptep/ptac097}{PTEP {\bf 2022} (2022)
  083C01}.

\bibitem{Gunion:1989we}
J.~F. Gunion, H.~E. Haber, G.~L. Kane, and S.~Dawson, {\em {The Higgs Hunter's
  Guide}}. Front. Phys. {\bf 80} (2000)  .

\bibitem{Aoki:2009ha}
M.~Aoki, S.~Kanemura, K.~Tsumura, and K.~Yagyu, {\em {Models of Yukawa
  interaction in the two Higgs doublet model, and their collider
  phenomenology}}. \href{http://dx.doi.org/10.1103/PhysRevD.80.015017}{Phys.
  Rev. D {\bf 80} (2009)  015017}, \href{http://arxiv.org/abs/0902.4665}{{\tt
  arXiv:0902.4665 [hep-ph]}}.

\bibitem{Branco:2011iw}
G.~C. Branco, P.~M. Ferreira, L.~Lavoura, M.~N. Rebelo, M.~Sher, and J.~P.
  Silva, {\em {Theory and phenomenology of two-Higgs-doublet models}}.
  \href{http://dx.doi.org/10.1016/j.physrep.2012.02.002}{Phys. Rept. {\bf 516}
  (2012)  1--102}, \href{http://arxiv.org/abs/1106.0034}{{\tt arXiv:1106.0034
  [hep-ph]}}.

\bibitem{Hessenberger:2016atw}
S.~Hessenberger and W.~Hollik, {\em {Two-loop corrections to the $\rho$
  parameter in Two-Higgs-Doublet Models}}.
  \href{http://dx.doi.org/10.1140/epjc/s10052-017-4734-8}{Eur. Phys. J. C {\bf
  77} (2017) no.~3, 178}, \href{http://arxiv.org/abs/1607.04610}{{\tt
  arXiv:1607.04610 [hep-ph]}}.

\bibitem{Hessenberger:2022tcx}
S.~Hessenberger and W.~Hollik, {\em {Two-loop improved predictions for $\mathbf
  {M_W}$ and $\mathbf {sin^2\theta _{eff}}$ in Two-Higgs-Doublet models}}.
  \href{http://dx.doi.org/10.1140/epjc/s10052-022-10933-6}{Eur. Phys. J. C {\bf
  82} (2022) no.~10, 970}, \href{http://arxiv.org/abs/2207.03845}{{\tt
  arXiv:2207.03845 [hep-ph]}}.

\bibitem{Braathen:2017izn}
J.~Braathen, M.~D. Goodsell, and F.~Staub, {\em {Supersymmetric and
  non-supersymmetric models without catastrophic Goldstone bosons}}.
  \href{http://dx.doi.org/10.1140/epjc/s10052-017-5303-x}{Eur. Phys. J. C {\bf
  77} (2017) no.~11, 757}, \href{http://arxiv.org/abs/1706.05372}{{\tt
  arXiv:1706.05372 [hep-ph]}}.

\bibitem{Braathen:2019pxr}
J.~Braathen and S.~Kanemura, {\em {On two-loop corrections to the Higgs
  trilinear coupling in models with extended scalar sectors}}.
  \href{http://dx.doi.org/10.1016/j.physletb.2019.07.021}{Phys. Lett. B {\bf
  796} (2019)  38--46}, \href{http://arxiv.org/abs/1903.05417}{{\tt
  arXiv:1903.05417 [hep-ph]}}.

\bibitem{Braathen:2019zoh}
J.~Braathen and S.~Kanemura, {\em {Leading two-loop corrections to the Higgs
  boson self-couplings in models with extended scalar sectors}}.
  \href{http://dx.doi.org/10.1140/epjc/s10052-020-7723-2}{Eur. Phys. J. C {\bf
  80} (2020) no.~3, 227}, \href{http://arxiv.org/abs/1911.11507}{{\tt
  arXiv:1911.11507 [hep-ph]}}.

\bibitem{Degrassi:2023eii}
G.~Degrassi and P.~Slavich, {\em {On the two-loop BSM corrections to
  $h\longrightarrow \gamma \gamma $ in the aligned THDM}}.
  \href{http://dx.doi.org/10.1140/epjc/s10052-023-12097-3}{Eur. Phys. J. C {\bf
  83} (2023) no.~10, 941}, \href{http://arxiv.org/abs/2307.02476}{{\tt
  arXiv:2307.02476 [hep-ph]}}.

\bibitem{Aiko:2023nqj}
M.~Aiko, J.~Braathen, and S.~Kanemura, {\em {Leading two-loop corrections to
  the Higgs di-photon decay in the Inert Doublet Model}}.
  \href{http://arxiv.org/abs/2307.14976}{{\tt arXiv:2307.14976 [hep-ph]}}.

\bibitem{Ross:1975fq}
D.~A. Ross and M.~J.~G. Veltman, {\em {Neutral Currents in Neutrino
  Experiments}}. \href{http://dx.doi.org/10.1016/0550-3213(75)90485-X}{Nucl.
  Phys. B {\bf 95} (1975)  135--147}.

\bibitem{Passarino:1989py}
G.~Passarino, {\em {The Interplay Between the Top Quark Mass and the Structure
  of the Higgs System}}.
  \href{http://dx.doi.org/10.1016/0370-2693(89)90694-1}{Phys. Lett. B {\bf 231}
  (1989)  458--462}.

\bibitem{Lynn:1990zk}
B.~W. Lynn and E.~Nardi, {\em {Radiative corrections in unconstrained SU(2) x
  U(1) and the top mass problem}}.
  \href{http://dx.doi.org/10.1016/0550-3213(92)90486-U}{Nucl. Phys. B {\bf 381}
  (1992)  467--500}.

\bibitem{Blank:1997qa}
T.~Blank and W.~Hollik, {\em {Precision observables in SU(2) x U(1) models with
  an additional Higgs triplet}}.
  \href{http://dx.doi.org/10.1016/S0550-3213(97)00785-2}{Nucl. Phys. B {\bf
  514} (1998)  113--134}, \href{http://arxiv.org/abs/hep-ph/9703392}{{\tt
  arXiv:hep-ph/9703392}}.

\bibitem{Forshaw:2001xq}
J.~R. Forshaw, D.~A. Ross, and B.~E. White, {\em {Higgs mass bounds in a
  triplet model}}. \href{http://dx.doi.org/10.1088/1126-6708/2001/10/007}{JHEP
  {\bf 10} (2001)  007}, \href{http://arxiv.org/abs/hep-ph/0107232}{{\tt
  arXiv:hep-ph/0107232}}.

\bibitem{Chen:2005jx}
M.-C. Chen, S.~Dawson, and T.~Krupovnickas, {\em {Constraining new models with
  precision electroweak data}}.
  \href{http://dx.doi.org/10.1142/S0217751X0603388X}{Int. J. Mod. Phys. A {\bf
  21} (2006)  4045--4070}, \href{http://arxiv.org/abs/hep-ph/0504286}{{\tt
  arXiv:hep-ph/0504286}}.

\bibitem{Chen:2006pb}
M.-C. Chen, S.~Dawson, and T.~Krupovnickas, {\em {Higgs triplets and limits
  from precision measurements}}.
  \href{http://dx.doi.org/10.1103/PhysRevD.74.035001}{Phys. Rev. D {\bf 74}
  (2006)  035001}, \href{http://arxiv.org/abs/hep-ph/0604102}{{\tt
  arXiv:hep-ph/0604102}}.

\bibitem{Chankowski:2006hs}
P.~H. Chankowski, S.~Pokorski, and J.~Wagner, {\em {(Non)decoupling of the
  Higgs triplet effects}}.
  \href{http://dx.doi.org/10.1140/epjc/s10052-007-0259-x}{Eur. Phys. J. C {\bf
  50} (2007)  919--933}, \href{http://arxiv.org/abs/hep-ph/0605302}{{\tt
  arXiv:hep-ph/0605302}}.

\bibitem{Chivukula:2007koj}
R.~S. Chivukula, N.~D. Christensen, and E.~H. Simmons, {\em {Low-energy
  effective theory, unitarity, and non-decoupling behavior in a model with
  heavy Higgs-triplet fields}}.
  \href{http://dx.doi.org/10.1103/PhysRevD.77.035001}{Phys. Rev. D {\bf 77}
  (2008)  035001}, \href{http://arxiv.org/abs/0712.0546}{{\tt arXiv:0712.0546
  [hep-ph]}}.

\bibitem{Chen:2008jg}
M.-C. Chen, S.~Dawson, and C.~B. Jackson, {\em {Higgs Triplets, Decoupling, and
  Precision Measurements}}.
  \href{http://dx.doi.org/10.1103/PhysRevD.78.093001}{Phys. Rev. D {\bf 78}
  (2008)  093001}, \href{http://arxiv.org/abs/0809.4185}{{\tt arXiv:0809.4185
  [hep-ph]}}.

\bibitem{FileviezPerez:2008bj}
P.~Fileviez~Perez, H.~H. Patel, M.~J. Ramsey-Musolf, and K.~Wang, {\em {Triplet
  Scalars and Dark Matter at the LHC}}.
  \href{http://dx.doi.org/10.1103/PhysRevD.79.055024}{Phys. Rev. D {\bf 79}
  (2009)  055024}, \href{http://arxiv.org/abs/0811.3957}{{\tt arXiv:0811.3957
  [hep-ph]}}.

\bibitem{Wang:2013jba}
L.~Wang and X.-F. Han, {\em {LHC diphoton and Z+photon Higgs signals in the
  Higgs triplet model with Y = 0}}.
  \href{http://dx.doi.org/10.1007/JHEP03(2014)010}{JHEP {\bf 03} (2014)  010},
  \href{http://arxiv.org/abs/1303.4490}{{\tt arXiv:1303.4490 [hep-ph]}}.

\bibitem{Chabab:2018ert}
M.~Chabab, M.~C. Peyran\`ere, and L.~Rahili, {\em {Probing the Higgs sector of
  $Y=0$ Higgs Triplet Model at LHC}}.
  \href{http://dx.doi.org/10.1140/epjc/s10052-018-6339-2}{Eur. Phys. J. C {\bf
  78} (2018) no.~10, 873}, \href{http://arxiv.org/abs/1805.00286}{{\tt
  arXiv:1805.00286 [hep-ph]}}.

\bibitem{Bell:2020gug}
N.~F. Bell, M.~J. Dolan, L.~S. Friedrich, M.~J. Ramsey-Musolf, and R.~R.
  Volkas, {\em {Two-Step Electroweak Symmetry-Breaking: Theory Meets
  Experiment}}. \href{http://dx.doi.org/10.1007/JHEP05(2020)050}{JHEP {\bf 05}
  (2020)  050}, \href{http://arxiv.org/abs/2001.05335}{{\tt arXiv:2001.05335
  [hep-ph]}}.

\bibitem{Chiang:2020rcv}
C.-W. Chiang, G.~Cottin, Y.~Du, K.~Fuyuto, and M.~J. Ramsey-Musolf, {\em
  {Collider Probes of Real Triplet Scalar Dark Matter}}.
  \href{http://dx.doi.org/10.1007/JHEP01(2021)198}{JHEP {\bf 01} (2021)  198},
  \href{http://arxiv.org/abs/2003.07867}{{\tt arXiv:2003.07867 [hep-ph]}}.

\bibitem{Ashanujjaman:2023etj}
S.~Ashanujjaman, S.~Banik, G.~Coloretti, A.~Crivellin, B.~Mellado, and A.-T.
  Mulaudzi, {\em {SU(2)L triplet scalar as the origin of the 95~GeV excess?}}
  \href{http://dx.doi.org/10.1103/PhysRevD.108.L091704}{Phys. Rev. D {\bf 108}
  (2023) no.~9, L091704}, \href{http://arxiv.org/abs/2306.15722}{{\tt
  arXiv:2306.15722 [hep-ph]}}.

\bibitem{Butterworth:2023rnw}
J.~Butterworth, H.~Debnath, P.~Fileviez~Perez, and F.~Mitchell, {\em {Custodial
  symmetry breaking and Higgs boson signatures at the LHC}}.
  \href{http://dx.doi.org/10.1103/PhysRevD.109.095014}{Phys. Rev. D {\bf 109}
  (2024) no.~9, 095014}, \href{http://arxiv.org/abs/2309.10027}{{\tt
  arXiv:2309.10027 [hep-ph]}}.

\bibitem{Ashanujjaman:2024pky}
S.~Ashanujjaman, S.~Banik, G.~Coloretti, A.~Crivellin, S.~P. Maharathy, and
  B.~Mellado, {\em {Explaining the $\gamma\gamma+X$ Excesses at $\approx$151.5
  GeV via the Drell-Yan Production of a Higgs Triplet}}.
  \href{http://arxiv.org/abs/2402.00101}{{\tt arXiv:2402.00101 [hep-ph]}}.

\bibitem{Crivellin:2024uhc}
A.~Crivellin, S.~Ashanujjaman, S.~Banik, G.~Coloretti, S.~P. Maharathy, and
  B.~Mellado, {\em {Growing Evidence for a Higgs Triplet}}.
  \href{http://arxiv.org/abs/2404.14492}{{\tt arXiv:2404.14492 [hep-ph]}}.

\bibitem{CDF:2022hxs}
{\bf CDF} Collaboration, T.~Aaltonen {\em et al.}, {\em {High-precision
  measurement of the $W$ boson mass with the CDF II detector}}.
  \href{http://dx.doi.org/10.1126/science.abk1781}{Science {\bf 376} (2022)
  no.~6589, 170--176}.

\bibitem{ATLAS:2024erm}
{\bf ATLAS} Collaboration, G.~Aad {\em et al.}, {\em {Measurement of the
  W-boson mass and width with the ATLAS detector using proton-proton collisions
  at $\sqrt{s}$ = 7 TeV}}. \href{http://arxiv.org/abs/2403.15085}{{\tt
  arXiv:2403.15085 [hep-ex]}}.

\bibitem{CMS:2024nau}
{\bf CMS} Collaboration, {\em {Measurement of the W boson mass in proton-proton
  collisions at sqrts = 13 TeV}}.
  \href{http://arxiv.org/abs/CMS-PAS-SMP-23-002}{{\tt CMS-PAS-SMP-23-002}}.

\bibitem{FileviezPerez:2022lxp}
P.~Fileviez~Perez, H.~H. Patel, and A.~D. Plascencia, {\em {On the W mass and
  new Higgs bosons}}.
  \href{http://dx.doi.org/10.1016/j.physletb.2022.137371}{Phys. Lett. B {\bf
  833} (2022)  137371}, \href{http://arxiv.org/abs/2204.07144}{{\tt
  arXiv:2204.07144 [hep-ph]}}.

\bibitem{Senjanovic:2022zwy}
G.~Senjanovi\'c and M.~Zantedeschi, {\em {SU(5) grand unification and W-boson
  mass}}. \href{http://dx.doi.org/10.1016/j.physletb.2022.137653}{Phys. Lett. B
  {\bf 837} (2023)  137653}, \href{http://arxiv.org/abs/2205.05022}{{\tt
  arXiv:2205.05022 [hep-ph]}}.

\bibitem{Shifman:1979eb}
M.~A. Shifman, A.~I. Vainshtein, M.~B. Voloshin, and V.~I. Zakharov, {\em
  {Low-Energy Theorems for Higgs Boson Couplings to Photons}}. Sov. J. Nucl.
  Phys. {\bf 30} (1979)  711--716.

\bibitem{Kniehl:1995tn}
B.~A. Kniehl and M.~Spira, {\em {Low-energy theorems in Higgs physics}}.
  \href{http://dx.doi.org/10.1007/s002880050007}{Z. Phys. C {\bf 69} (1995)
  77--88}, \href{http://arxiv.org/abs/hep-ph/9505225}{{\tt
  arXiv:hep-ph/9505225}}.

\bibitem{Staub:2008uz}
F.~Staub, {\em {SARAH}}. \href{http://arxiv.org/abs/0806.0538}{{\tt
  arXiv:0806.0538 [hep-ph]}}.

\bibitem{Staub:2009bi}
F.~Staub, {\em {From Superpotential to Model Files for FeynArts and
  CalcHep/CompHep}}. \href{http://dx.doi.org/10.1016/j.cpc.2010.01.011}{Comput.
  Phys. Commun. {\bf 181} (2010)  1077--1086},
  \href{http://arxiv.org/abs/0909.2863}{{\tt arXiv:0909.2863 [hep-ph]}}.

\bibitem{Staub:2010jh}
F.~Staub, {\em {Automatic Calculation of supersymmetric Renormalization Group
  Equations and Self Energies}}.
  \href{http://dx.doi.org/10.1016/j.cpc.2010.11.030}{Comput. Phys. Commun. {\bf
  182} (2011)  808--833}, \href{http://arxiv.org/abs/1002.0840}{{\tt
  arXiv:1002.0840 [hep-ph]}}.

\bibitem{Staub:2012pb}
F.~Staub, {\em {SARAH 3.2: Dirac Gauginos, UFO output, and more}}.
  \href{http://dx.doi.org/10.1016/j.cpc.2013.02.019}{Comput. Phys. Commun. {\bf
  184} (2013)  1792--1809}, \href{http://arxiv.org/abs/1207.0906}{{\tt
  arXiv:1207.0906 [hep-ph]}}.

\bibitem{Staub:2013tta}
F.~Staub, {\em {SARAH 4 : A tool for (not only SUSY) model builders}}.
  \href{http://dx.doi.org/10.1016/j.cpc.2014.02.018}{Comput. Phys. Commun. {\bf
  185} (2014)  1773--1790}, \href{http://arxiv.org/abs/1309.7223}{{\tt
  arXiv:1309.7223 [hep-ph]}}.

\bibitem{Chardonnet:1993wd}
P.~Chardonnet, P.~Salati, and P.~Fayet, {\em {Heavy triplet neutrinos as a new
  dark matter option}}.
  \href{http://dx.doi.org/10.1016/0550-3213(93)90101-T}{Nucl. Phys. B {\bf 394}
  (1993)  35--72}.

\bibitem{Benakli:2022gjn}
K.~Benakli, M.~Goodsell, W.~Ke, and P.~Slavich, {\em {W boson mass in minimal
  Dirac gaugino scenarios}}.
  \href{http://dx.doi.org/10.1140/epjc/s10052-022-11132-z}{Eur. Phys. J. C {\bf
  83} (2023) no.~1, 43}, \href{http://arxiv.org/abs/2208.05867}{{\tt
  arXiv:2208.05867 [hep-ph]}}.

\bibitem{Cirelli:2005uq}
M.~Cirelli, N.~Fornengo, and A.~Strumia, {\em {Minimal dark matter}}.
  \href{http://dx.doi.org/10.1016/j.nuclphysb.2006.07.012}{Nucl. Phys. B {\bf
  753} (2006)  178--194}, \href{http://arxiv.org/abs/hep-ph/0512090}{{\tt
  arXiv:hep-ph/0512090}}.

\bibitem{Khan:2016sxm}
N.~Khan, {\em {Exploring the hyperchargeless Higgs triplet model up to the
  Planck scale}}. \href{http://dx.doi.org/10.1140/epjc/s10052-018-5766-4}{Eur.
  Phys. J. C {\bf 78} (2018) no.~4, 341},
  \href{http://arxiv.org/abs/1610.03178}{{\tt arXiv:1610.03178 [hep-ph]}}.

\bibitem{Forshaw:2003kh}
J.~R. Forshaw, A.~Sabio~Vera, and B.~E. White, {\em {Mass bounds in a model
  with a triplet Higgs}}.
  \href{http://dx.doi.org/10.1088/1126-6708/2003/06/059}{JHEP {\bf 06} (2003)
  059}, \href{http://arxiv.org/abs/hep-ph/0302256}{{\tt arXiv:hep-ph/0302256}}.

\bibitem{Hahn:2000kx}
T.~Hahn, {\em {Generating Feynman diagrams and amplitudes with FeynArts 3}}.
  \href{http://dx.doi.org/10.1016/S0010-4655(01)00290-9}{Comput. Phys. Commun.
  {\bf 140} (2001)  418--431}, \href{http://arxiv.org/abs/hep-ph/0012260}{{\tt
  arXiv:hep-ph/0012260}}.

\bibitem{Davydychev:1992mt}
A.~I. Davydychev and J.~B. Tausk, {\em {Two loop selfenergy diagrams with
  different masses and the momentum expansion}}.
  \href{http://dx.doi.org/10.1016/0550-3213(93)90338-P}{Nucl. Phys. B {\bf 397}
  (1993)  123--142}.

\bibitem{Zheng:1990qa}
H.-Q. Zheng and D.-D. Wu, {\em {First order QCD corrections to the decay of the
  Higgs boson into two photons}}.
  \href{http://dx.doi.org/10.1103/PhysRevD.42.3760}{Phys. Rev. D {\bf 42}
  (1990)  3760--3763}.

\bibitem{Djouadi:1990aj}
A.~Djouadi, M.~Spira, J.~J. van~der Bij, and P.~M. Zerwas, {\em {QCD
  corrections to $\gamma\gamma$ decays of Higgs particles in the intermediate
  mass range}}. \href{http://dx.doi.org/10.1016/0370-2693(91)90879-U}{Phys.
  Lett. B {\bf 257} (1991)  187--190}.

\bibitem{Dawson:1992cy}
S.~Dawson and R.~P. Kauffman, {\em {QCD corrections to H ---\ensuremath{>}
  $\gamma\gamma$}}. \href{http://dx.doi.org/10.1103/PhysRevD.47.1264}{Phys.
  Rev. D {\bf 47} (1993)  1264--1267}.

\bibitem{Melnikov:1993tj}
K.~Melnikov and O.~I. Yakovlev, {\em {Higgs ---\ensuremath{>} two photon decay:
  QCD radiative correction}}.
  \href{http://dx.doi.org/10.1016/0370-2693(93)90507-E}{Phys. Lett. B {\bf 312}
  (1993)  179--183}, \href{http://arxiv.org/abs/hep-ph/9302281}{{\tt
  arXiv:hep-ph/9302281}}.

\bibitem{Djouadi:1993ji}
A.~Djouadi, M.~Spira, and P.~M. Zerwas, {\em {Two photon decay widths of Higgs
  particles}}. \href{http://dx.doi.org/10.1016/0370-2693(93)90564-X}{Phys.
  Lett. B {\bf 311} (1993)  255--260},
  \href{http://arxiv.org/abs/hep-ph/9305335}{{\tt arXiv:hep-ph/9305335}}.

\bibitem{Inoue:1994jq}
M.~Inoue, R.~Najima, T.~Oka, and J.~Saito, {\em {QCD corrections to two photon
  decay of the Higgs boson and its reverse process}}.
  \href{http://dx.doi.org/10.1142/S0217732394001003}{Mod. Phys. Lett. A {\bf 9}
  (1994)  1189--1194}.

\bibitem{Fleischer:2004vb}
J.~Fleischer, O.~V. Tarasov, and V.~O. Tarasov, {\em {Analytical result for the
  two loop QCD correction to the decay H ---\ensuremath{>} 2 gamma}}.
  \href{http://dx.doi.org/10.1016/j.physletb.2004.01.063}{Phys. Lett. B {\bf
  584} (2004)  294--297}, \href{http://arxiv.org/abs/hep-ph/0401090}{{\tt
  arXiv:hep-ph/0401090}}.

\bibitem{Liao:1996td}
Y.~Liao and X.-y. Li, {\em {O (alpha**2 G(F)m(t)**2) contributions to H
  ---\ensuremath{>} $\gamma\gamma$}}.
  \href{http://dx.doi.org/10.1016/S0370-2693(97)00089-0}{Phys. Lett. B {\bf
  396} (1997)  225--230}, \href{http://arxiv.org/abs/hep-ph/9605310}{{\tt
  arXiv:hep-ph/9605310}}.

\bibitem{Djouadi:1997rj}
A.~Djouadi, P.~Gambino, and B.~A. Kniehl, {\em {Two loop electroweak heavy
  fermion corrections to Higgs boson production and decay}}.
  \href{http://dx.doi.org/10.1016/S0550-3213(98)00147-3}{Nucl. Phys. B {\bf
  523} (1998)  17--39}, \href{http://arxiv.org/abs/hep-ph/9712330}{{\tt
  arXiv:hep-ph/9712330}}.

\bibitem{Fugel:2004ug}
F.~Fugel, B.~A. Kniehl, and M.~Steinhauser, {\em {Two loop electroweak
  correction of O(G(F)M(t)**2) to the Higgs-boson decay into photons}}.
  \href{http://dx.doi.org/10.1016/j.nuclphysb.2004.09.018}{Nucl. Phys. B {\bf
  702} (2004)  333--345}, \href{http://arxiv.org/abs/hep-ph/0405232}{{\tt
  arXiv:hep-ph/0405232}}.

\bibitem{Degrassi:2005mc}
G.~Degrassi and F.~Maltoni, {\em {Two-loop electroweak corrections to the
  Higgs-boson decay H ---\ensuremath{>} $\gamma\gamma$}}.
  \href{http://dx.doi.org/10.1016/j.nuclphysb.2005.06.027}{Nucl. Phys. B {\bf
  724} (2005)  183--196}, \href{http://arxiv.org/abs/hep-ph/0504137}{{\tt
  arXiv:hep-ph/0504137}}.

\bibitem{Aglietti:2004nj}
U.~Aglietti, R.~Bonciani, G.~Degrassi, and A.~Vicini, {\em {Two loop light
  fermion contribution to Higgs production and decays}}.
  \href{http://dx.doi.org/10.1016/j.physletb.2004.06.063}{Phys. Lett. B {\bf
  595} (2004)  432--441}, \href{http://arxiv.org/abs/hep-ph/0404071}{{\tt
  arXiv:hep-ph/0404071}}.

\bibitem{Aglietti:2004ki}
U.~Aglietti, R.~Bonciani, G.~Degrassi, and A.~Vicini, {\em {Master integrals
  for the two-loop light fermion contributions to gg ---\ensuremath{>} H and H
  ---\ensuremath{>} $\gamma\gamma$}}.
  \href{http://dx.doi.org/10.1016/j.physletb.2004.09.001}{Phys. Lett. B {\bf
  600} (2004)  57--64}, \href{http://arxiv.org/abs/hep-ph/0407162}{{\tt
  arXiv:hep-ph/0407162}}.

\bibitem{Passarino:2007fp}
G.~Passarino, C.~Sturm, and S.~Uccirati, {\em {Complete Two-Loop Corrections to
  H ---\ensuremath{>} $\gamma\gamma$}}.
  \href{http://dx.doi.org/10.1016/j.physletb.2007.09.002}{Phys. Lett. B {\bf
  655} (2007)  298--306}, \href{http://arxiv.org/abs/0707.1401}{{\tt
  arXiv:0707.1401 [hep-ph]}}.

\bibitem{Actis:2008ts}
S.~Actis, G.~Passarino, C.~Sturm, and S.~Uccirati, {\em {NNLO Computational
  Techniques: The Cases H ---\ensuremath{>} $\gamma\gamma$ and H
  ---\ensuremath{>} g g}}.
  \href{http://dx.doi.org/10.1016/j.nuclphysb.2008.11.024}{Nucl. Phys. B {\bf
  811} (2009)  182--273}, \href{http://arxiv.org/abs/0809.3667}{{\tt
  arXiv:0809.3667 [hep-ph]}}.

\bibitem{CMS:2022dwd}
{\bf CMS} Collaboration, A.~Tumasyan {\em et al.}, {\em {A portrait of the
  Higgs boson by the CMS experiment ten years after the discovery.}}
  \href{http://dx.doi.org/10.1038/s41586-022-04892-x}{Nature {\bf 607} (2022)
  no.~7917, 60--68}, \href{http://arxiv.org/abs/2207.00043}{{\tt
  arXiv:2207.00043 [hep-ex]}}.

\bibitem{Cepeda:2019klc}
M.~Cepeda {\em et al.}, {\em {Report from Working Group 2}: {Higgs Physics at
  the HL-LHC and HE-LHC}}.
  \href{http://dx.doi.org/10.23731/CYRM-2019-007.221}{CERN Yellow Rep. Monogr.
  {\bf 7} (2019)  221--584}, \href{http://arxiv.org/abs/1902.00134}{{\tt
  arXiv:1902.00134 [hep-ph]}}.

\bibitem{Braathen:2016cqe}
J.~Braathen and M.~D. Goodsell, {\em {Avoiding the Goldstone Boson Catastrophe
  in general renormalisable field theories at two loops}}.
  \href{http://dx.doi.org/10.1007/JHEP12(2016)056}{JHEP {\bf 12} (2016)  056},
  \href{http://arxiv.org/abs/1609.06977}{{\tt arXiv:1609.06977 [hep-ph]}}.

\bibitem{Braathen:2018htl}
J.~Braathen, M.~D. Goodsell, and P.~Slavich, {\em {Matching renormalisable
  couplings: simple schemes and a plot}}.
  \href{http://dx.doi.org/10.1140/epjc/s10052-019-7093-9}{Eur. Phys. J. C {\bf
  79} (2019) no.~8, 669}, \href{http://arxiv.org/abs/1810.09388}{{\tt
  arXiv:1810.09388 [hep-ph]}}.

\end{thebibliography}\endgroup

\end{document}